\documentclass{article}
\pdfoutput=1
\usepackage{amsmath,amsfonts,amssymb}
\usepackage{epsfig}
\usepackage{cite, bbm, color}
\newcommand{\eq}[1]{\begin{equation}\label{#1}}
\newcommand{\be}{\begin{equation}}
\newcommand{\ee}{\end{equation}}
\newcommand{\benn}{\nonumber\begin{equation}}
\newcommand{\eenn}{\nonumber\end{equation}}
\newcommand{\dks}{D_\text{KS}}
\def\bea{\begin{eqnarray}} \def\eea{\end{eqnarray}}
\def\beann{\begin{eqnarray*}} \def\eeann{\end{eqnarray*}}
\def\lsim{\raise0.3ex\hbox{$<$\kern-0.75em\raise-1.1ex\hbox{$\sim$}}}
\def\gsim{\raise0.3ex\hbox{$>$\kern-0.75em\raise-1.1ex\hbox{$\sim$}}}

\newcommand{\en}{\end{equation}}
\newcommand{\ba}{\begin{eqnarray}}
\newcommand{\ea}{\end{eqnarray}}

\newcommand{\Dov}{D_{\mbox{\tiny{ov}}}}

\newcommand{\ide}{\mathbf{1}}
\newcommand{\ear}[1]{\begin{eqnarray}\label{#1}}
\newcommand{\enar}{\end{eqnarray}}

\renewcommand{\vec}[1]{{\bf #1}}

\def\openone{\rlap 1\kern 0.22ex 1}
\newcommand{\bi}{\begin{itemize}}
\newcommand{\ei}{\end{itemize}}

\begin{document}

\begin{titlepage}
\begin{flushright} 
                CERN-PH-TH/2012-040\\
                HIP-2012-06/TH\\
                NSF-KITP-12-017\\
                YITP-12-7
\end{flushright}
\vspace*{0.5cm}
\begin{center}
{\Large\bf Numerical properties of staggered quarks \\ with a taste-dependent mass term}
\end{center}

\vskip1.3cm

\centerline{Philippe~de~Forcrand,$^{a,b,c,d}$ Aleksi~Kurkela$^{e}$ and  Marco~Panero$^{c,f}$}
\vskip1.5cm
\centerline{\sl  $^a$ Institute for Theoretical Physics,  ETH Z\"urich, CH-8093 Z\"urich, Switzerland}
\vskip0.5cm
\centerline{\sl  $^b$ CERN, Physics Department, TH Unit, CH-1211 Gen\`eve 23, Switzerland}

\vskip0.5cm

\centerline{\sl  $^c$ Kavli Institute for Theoretical Physics, University of California, Santa Barbara, CA 93106, USA}

\vskip0.5cm

\centerline{\sl  $^d$ Yukawa Institute for Theoretical Physics, Kyoto University, Kyoto 606-8502, Japan}

\vskip0.5cm

\centerline{\sl  $^e$ Department of Physics, McGill University, 3600 rue University, Montr\'eal, QC H3A 2T8, Canada}

\vskip0.5cm

\centerline{\sl  $^f$ Department of Physics and Helsinki Institute of Physics, University of Helsinki, FIN-00014 Helsinki, Finland}

\vskip0.5cm

\begin{center}

{\sl  e-mail:} \hskip 6mm \texttt{forcrand@phys.ethz.ch, aleksi.kurkela@mcgill.ca, marco.panero@helsinki.fi}

\end{center}

\vskip1.0cm

\begin{abstract}

The numerical properties of staggered Dirac operators with a 
taste-dependent mass term proposed by
Adams~\cite{staggeredoverlap1,staggeredoverlap2} and by 
Hoelbling~\cite{Hoelbling:2010jw} are compared with those of 
ordinary staggered and Wilson Dirac operators. In the free limit 
and on (quenched) interacting configurations, we consider their 
topological properties, their spectrum, and the resulting pion
mass. Although we also consider the spectral structure, 
topological properties, locality, and computational cost of an 
overlap operator with a staggered kernel, we call attention to 
the possibility of using the Adams and Hoelbling operators 
without the overlap construction. In particular, the Hoelbling
operator could be used to simulate two degenerate flavors without 
additive mass renormalization, and thus without fine-tuning in 
the chiral limit.

\end{abstract}

\end{titlepage}

\section{Introduction}

\label{sec:intro}
The spontaneous breakdown of chiral symmetry plays a central role
in the spectrum of light hadrons. Since it is an intrinsically 
non-perturbative phenomenon, the only way to study it from the 
first principles of QCD is via the lattice regularization. Yet, 
already many years ago Nielsen and Ninomiya proved that a 
translationally invariant, local lattice formulation of the QCD 
Dirac operator $D$, retaining chiral symmetry in the massless 
limit, and with the correct number of physical fermionic degrees 
of freedom, is forbidden~\cite{Nielsen_Ninomiya}. This no-go 
theorem can be circumvented, by constructing lattice fermions 
satisfying a \emph{modified} form of chiral 
symmetry~\cite{latticechiralsymmetry}, and obeying the Ginsparg-
Wilson relation~\cite{Ginsparg:1981bj}. Although explicit 
formulations of lattice Ginsparg-Wilson fermions are 
known~\cite{overlap_DW_perfect}, currently their practical use in 
realistic, large-scale lattice QCD simulations is still limited, 
due to the high computational overhead.

The most widely-used lattice discretizations of the Dirac 
operator are either based on the addition of a second-derivative 
term to the kinetic part of the quark 
action~\cite{Wilson_fermions} to remove (or ``quench'') the 
unphysical ``doubler'' modes in the continuum limit by giving 
them a mass ${\cal O}(a^{-1})$, or on a site-dependent spin 
diagonalization, which leads to the so-called staggered 
formulation~\cite{Kogut:1974ag}. The former approach introduces 
an explicit breaking of chiral symmetry, and, as a consequence, 
an additive renormalization of the quark mass, which has to be 
fine-tuned. In contrast, the staggered operator preserves a 
remnant of chiral symmetry (sufficient to forbid additive mass 
renormalization), and leads to a reduction of the matrix size. 
However, the staggered formulation only removes part of the 
unphysical modes, reducing the number of quark species in four 
($d$) spacetime dimensions from 16 ($2^d$) down to four 
($2^{d/2}$) ``tastes'', which become 
degenerate~\cite{Follana:2005km} (and consistent with the 
properties related to the global symmetries of the continuum 
Dirac operator~\cite{Bruckmann:2008xr}) in the $a \to 0$ limit. 
In order to simulate QCD with two light fermions, one then has to 
apply the so-called ``rooting trick'', which has been a subject 
of debate for the last few years~\cite{rooting_saga}.

Some recent works have discussed the idea of using a staggered 
kernel with a \emph{taste-dependent} mass term to obtain two (or 
one~\cite{deForcrand:2011ak, Hoelbling:2010jw}) massless fermion 
species. Such formulation, which is one of the various approaches 
aiming at minimally doubled 
fermions~\cite{minimallydoubledfermions}, could combine the 
advantages of the overlap construction with the computational 
efficiency of a staggered kernel. Furthermore, this 
formulation appears to be particularly attractive from the point 
of view of topological properties~\cite{staggeredoverlap1}.

Using a staggered operator with a ``flavored'' mass term as the 
kernel in an overlap construction is a very appealing idea, but 
the properties of such operators (with various taste-dependent 
mass terms) are interesting on their own. In fact, while the 
overlap construction completely removes the need for fine tuning 
to achieve massless fermions, it still leads to a considerable 
computational overhead. In contrast, using a staggered operator 
with taste-dependent mass \emph{\`a la} Wilson requires fine 
tuning to obtain exactly massless modes, but, by 
virtue of the reduced size of the operator, may still be a 
computationally competitive alternative to the usual Wilson 
discretization, while avoiding the rooting prescription.

This motivation led us to address a numerical investigation of 
different operators of this type that we present here 
(preliminary results have appeared in \cite{deForcrand:2011ak}). 
In the following, we present a systematic classification of the 
possible taste-dependent mass terms, discuss their analytical 
features in the free theory, and then move on to the interacting 
case that we study via numerical simulations. We perform an 
elementary measurement of the pion mass on a set of quenched 
configurations, and verify the expected PCAC behaviour as one 
approaches the chiral limit. In an Appendix, we also explore the 
properties of the staggered overlap operator proposed in 
\cite{staggeredoverlap1}, in comparison with the usual overlap 
based on the Wilson kernel. In particular, we compare the 
locality of the operators, and the computational cost of applying 
them to a vector and of solving for the quark propagator.

The structure of this paper is as follows. First, in 
sec.~\ref{sec:theoretical} we recall theoretical aspects of 
the construction of taste-dependent mass terms, and discuss their 
spectral structure in the free field case. Then, we address the 
interacting case, presenting our numerical studies in 
sec.~\ref{sec:numerical}. We summarize our findings and 
discuss their implications for possible future, large-scale 
applications of these operators in sec.~\ref{sec:conclusions}. Finally, 
in the appendix~\ref{appendix}, we report on our study of an overlap 
operator based on a staggered kernel, as proposed in ref.~\cite{staggeredoverlap1}.

\section{Theoretical formulation and general features}

\label{sec:theoretical}

The staggered operator~\cite{Kogut:1974ag}
\begin{equation}
\label{dks}
\dks = \frac{1}{2a} \sum_{\mu=1}^{d} \eta_\mu \left(   V_\mu - V_\mu^\dagger \right)
\end{equation}
with $\eta_\mu(x)=(-1)^{\sum_{\nu<\mu} x_\nu}$ and $\left( V_\mu \right)_{x,y}
= U_\mu(x) \delta_{x+a\hat\mu, y}$, is a computationally very 
efficient way to discretize the massless QCD Dirac operator on a 
$d$-dimensional Euclidean hypercubic lattice of spacing $a$. This 
operator is invariant under a global $U(1)$ symmetry, which can 
be interpreted as a remnant of chiral symmetry: in fact, $\dks$ 
anticommutes with the operator $\Gamma_{55}$ defined by $\left( 
\Gamma_{55}\right)_{x,y}=(-1)^{\sum_{\nu=1}^d x_\nu}$. In the 
free theory, one can easily see that in four dimensions the 
operator $\Gamma_{55}$ has $\gamma_5 \otimes \gamma_5$ structure 
in spin-taste space~\cite{spinflavorinterpretation}. The 
construction of $\dks$ is based on a local spin diagonalization, 
which, for the four-dimensional case, allows one to reduce the 
number of fermion components by a factor of $4$ with respect to the 
naive operator, and yields four tastes in the continuum limit. 
The degeneracy between these four tastes is explicitly broken by 
gauge interactions at finite lattice spacing $a$, but is 
recovered in the continuum limit $a \to 0$.

Recently, various works explored the idea of using staggered 
operators with taste-dependent 
mass terms~\cite{staggeredoverlap1, deForcrand:2011ak, Hoelbling:2010jw}. Following, e.g., the discussion in the classic 
paper by Golterman and Smit~\cite{Golterman:1984cy}, the possible 
matrix structures (in taste space) for a mass term can be 
classified as
\begin{itemize}
\item $\ide$ (``0-link''),
of the form $\delta_{x,y}$
\item $\gamma_\alpha$ (``1-link''),
involving a sum of terms, each containing 1 link $U_\mu$
\item $\sigma_{\alpha\beta}$ (``2-link''),
involving a sum of terms, each containing 2 links $U_\mu U_\nu$
\item $\gamma_5 \gamma_\alpha$ (``3-link''),
involving a sum of terms, each containing 3 links $U_\mu U_\nu U_\rho$
\item $\gamma_5$ (``4-link''),
involving a sum of terms, each containing 4 links $U_\mu U_\nu U_\rho U_\sigma$
\end{itemize}

It is highly desirable to preserve the symmetry 
$\Gamma_{55} D \Gamma_{55} = D^\dagger$, because it guarantees 
that $\det D$ is real, and non-negative (in the absence of real negative eigenvalues), thus avoiding a ``sign 
problem'' in the measure~\cite{signproblem}. This symmetry is 
satisfied only if the hermitian mass term connects sites of the 
same parity. Thus, we do not consider the 1-link or a 3-link mass 
terms further.

This leaves three possibilities: 0-, 2- and 4-link mass 
terms.\footnote{It is also possible to consider the above matrix 
possibilities with an extra factor 
$\Gamma_{55}$~\cite{Golterman:1984cy}. In that case, $\gamma_5$-
hermitian mass terms are obtained in the 0-, 1- and 3-link cases. 
However, we did not find a continuum-like dispersion relation for 
the real modes in any of these cases.}
The 0-link mass term corresponds to the usual staggered operator, 
with a taste-independent bare mass
\begin{equation}
\label{plainvanillastaggeredmass}
D_0 = \dks + m .
\end{equation}
The staggered operator with a 2-link mass term, which was 
discussed in refs.~\cite{deForcrand:2011ak, Hoelbling:2010jw}, 
can be written in the form
\begin{equation}
\label{twolinkkernel}
D_2 = \dks + \frac{\rho}{\sqrt{3}}\left(M_{12}+M_{13}+M_{14}+M_{23}-M_{24}+M_{34}\right),
\end{equation}
where (following the notation of ref.~\cite{Hoelbling:2010jw})
\begin{equation}
M_{\mu\nu}=i \eta_{\mu \nu} C_{\mu\nu},
\end{equation}
\begin{equation}
(\eta_{\mu\nu})_{x,y}=-(\eta_{\nu\mu})_{x,y}=(-1)^{\sum_{i=\mu+1}^{\nu} x_i}\delta_{x,y}\textrm{, for $\mu < \nu$},
\end{equation}
\begin{equation}
C_{\mu\nu}=\frac{1}{2}\left(C_\mu C_\nu+C_\nu C_\mu\right),
\end{equation}
\begin{equation}
\label{Cmudefinition}
C_\mu = \frac{1}{2} \left( V_\mu + V_\mu^\dagger \right).
\end{equation}
Finally, a staggered operator featuring a mass term with 
$\gamma_5$ structure in taste space~\cite{staggeredoverlap1} can 
be written as
\begin{equation}
\label{fourlinkkernel}
D_4 = \dks - \frac{\rho}{a}\Gamma_{55}\Gamma_{5},
\end{equation}
with
\begin{equation}
\Gamma_{5}=\eta_5 C,
\end{equation}
where
\begin{equation}
\label{eta5_definition}
\eta_5 (x) =\prod_{\mu=1}^4 \eta_\mu (x),
\end{equation}
while $C$ is the average of four-link parallel transporters 
joining sites at opposite corners of the elementary lattice 
hypercubes
\begin{equation}
\label{C_definition}
C = \frac{1}{4!} \sum_{\mbox{\tiny{perm}}} C_\mu C_\nu C_\rho C_\sigma.
\end{equation}
Note that the mass term appearing on the r.h.s. of 
eq.~(\ref{fourlinkkernel}) is Hermitean and commutes with 
$\Gamma_{55}$. 

\begin{figure}
\centerline{
\hspace*{-3.0cm}
\includegraphics[width=0.73\textwidth]{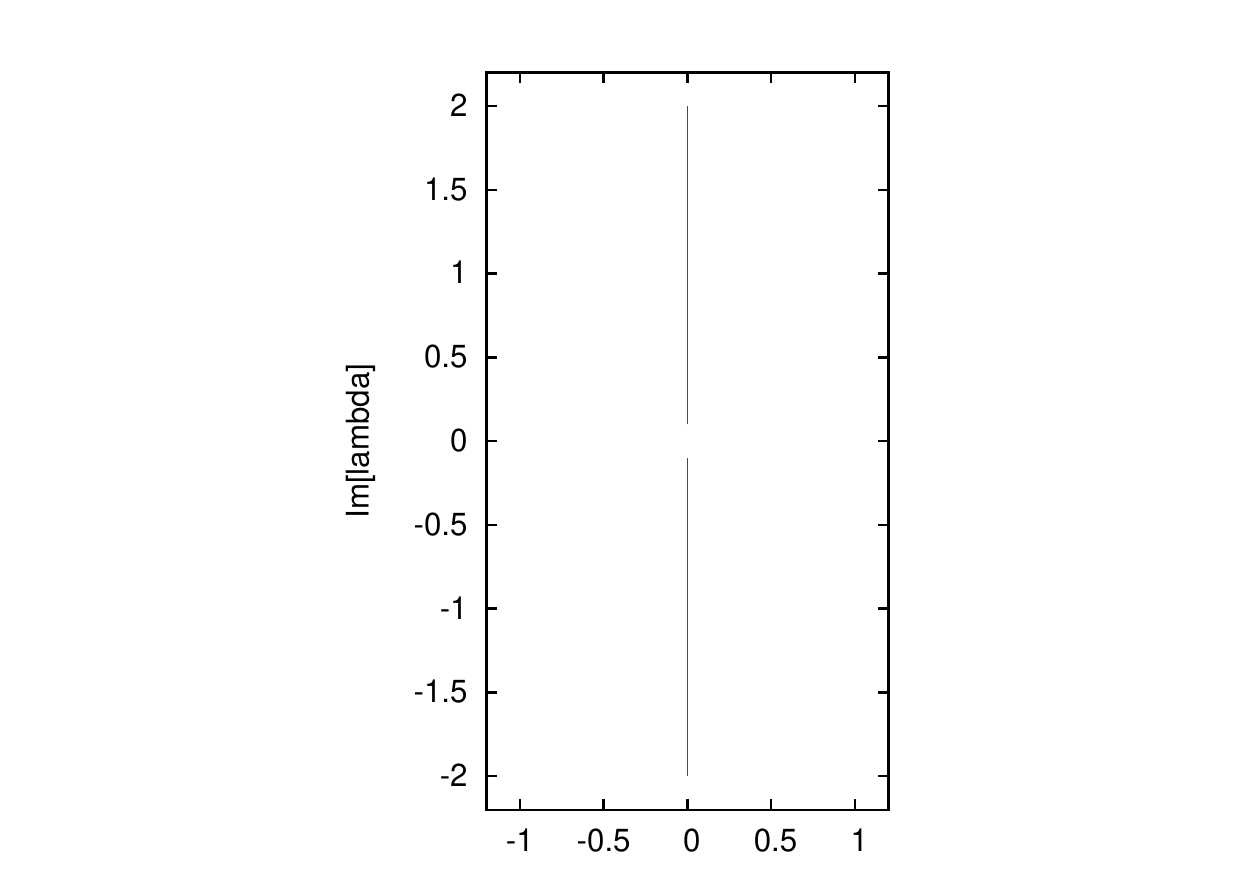}
\hspace*{-1.9cm}
\includegraphics[width=0.43\textwidth,viewport=-1 -10 495 402]{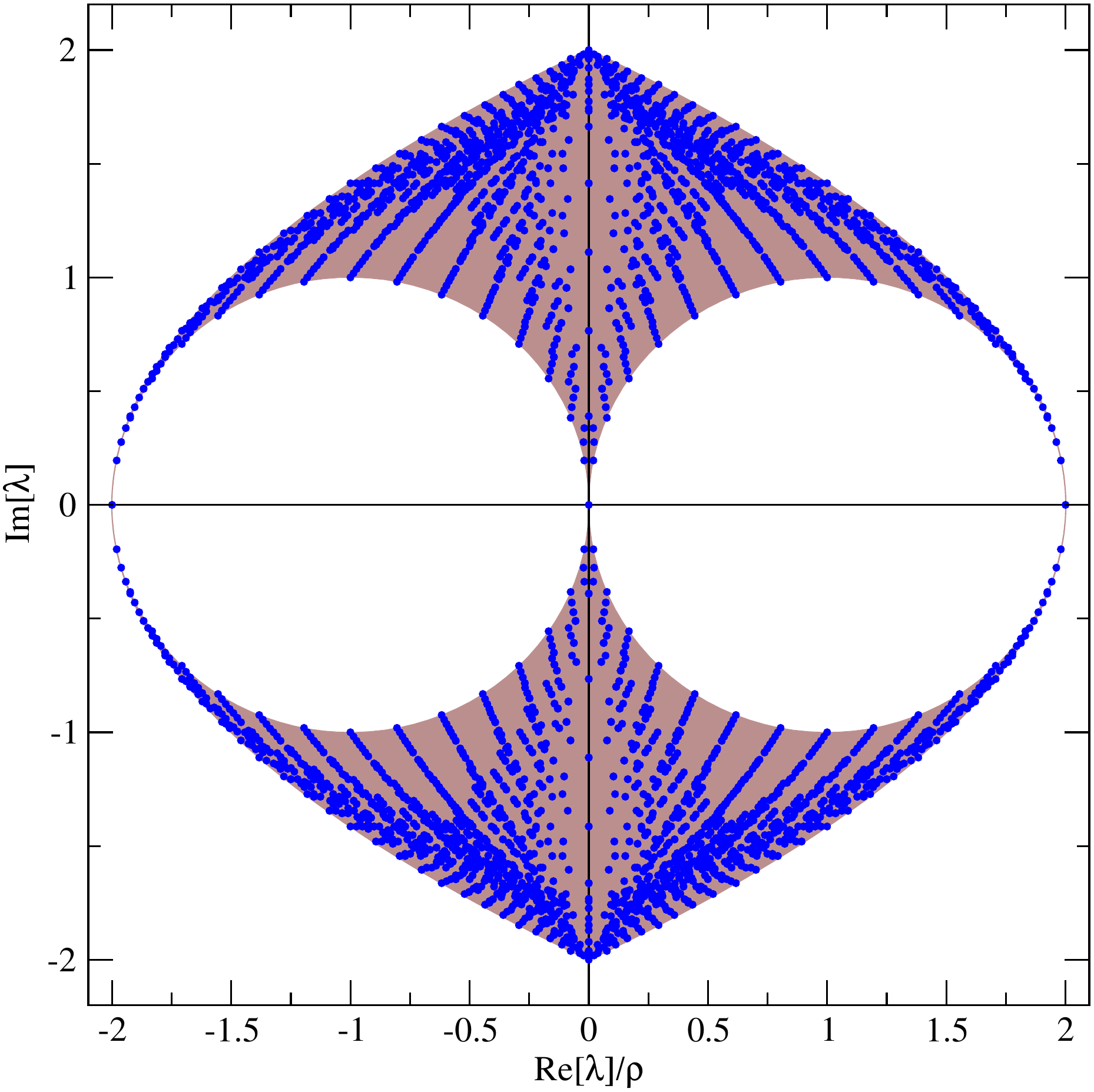}
\hspace*{0.5cm}
\includegraphics[width=0.295\textwidth,clip=true]{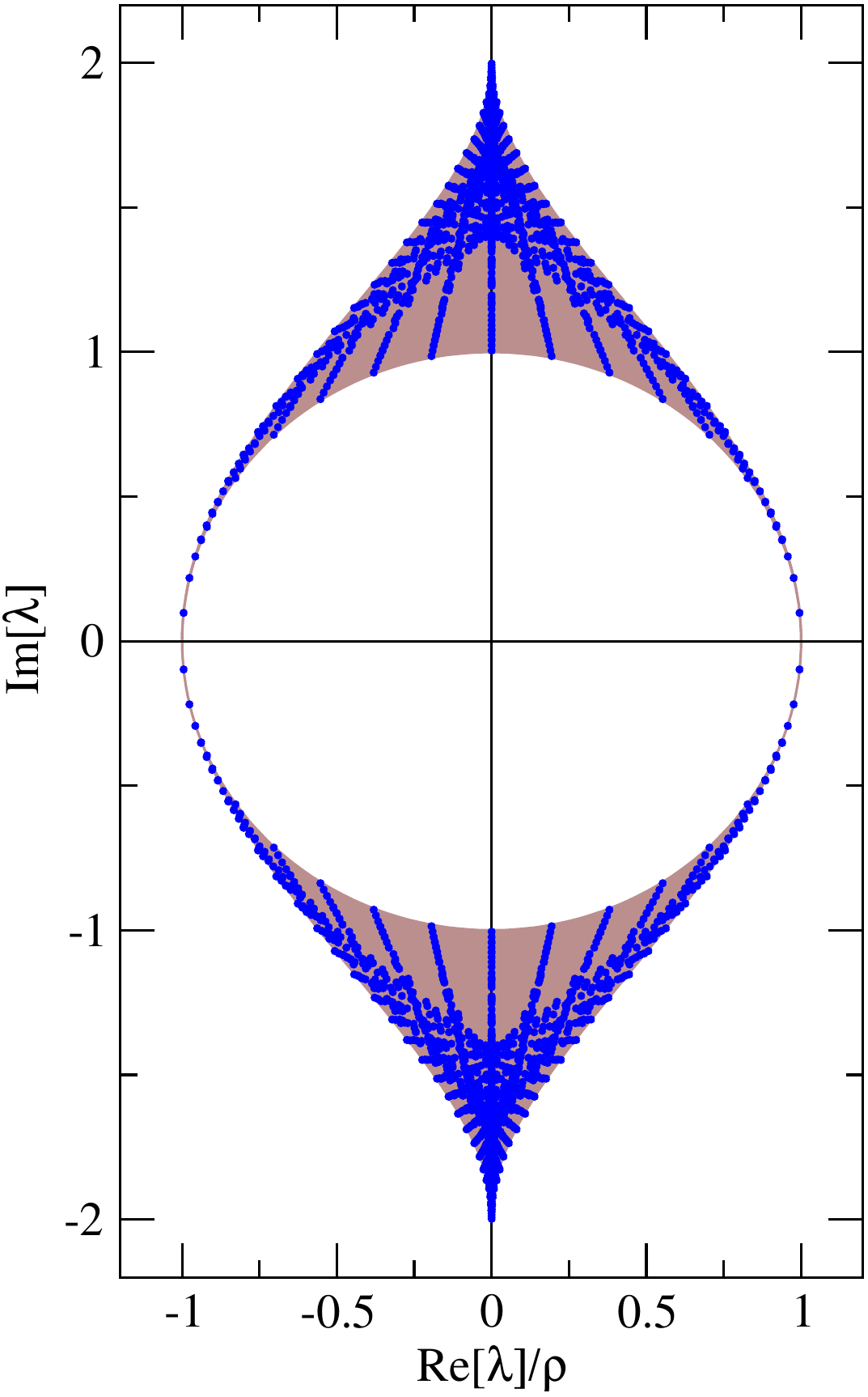}
}
\caption{Left panel: Spectrum of $\dks$ in the free limit. 
Central panel: Free spectrum of operator $D_2$ (eq.(\ref{twolinkkernel})), which includes a 
taste-dependent mass term with tensor-like structure in taste 
space (i.e., a 2-link mass term). Right panel: Free spectrum of
operator $D_4$ (eq.(\ref{fourlinkkernel})), which includes a taste-dependent mass term with 
$\gamma_5$ structure in taste space (i.e., a 4-link mass term).}
\label{fig:freespectra}
\end{figure}

To understand the properties of these three different types of 
operators it is instructive to start by discussing their spectra 
in the free limit. The three panels in Fig.~\ref{fig:freespectra} 
show the structure of the spectrum of eigenvalues for $\dks$ (for 
$D_0$, the spectrum is just trivially shifted by $m$), for $D_2$, 
and for $D_4$ in the non-interacting case.

In the free limit the eigenvalues of $D_0$ on a 
lattice with $N_\mu$ sites along the $\mu$ direction read
\begin{equation}
\label{D0_free_spectrum}
\lambda = m \pm i \sqrt{\sum_{i=1}^d \sin^2 p_\mu }, \qquad \mbox{with: }\;\; p_\mu=\frac{2\pi}{N_\mu} ( k_\mu + \varepsilon_\mu) , \qquad k_\mu \in \{ 0, 1, 2, \dots , L_\mu/2-1 \},
\end{equation}
with eight degenerate eigenvalues of both signs, and where $\varepsilon_\mu=0$ ($1/2$) if the fermionic field satisfies 
(anti-)periodic boundary conditions along the $\mu$ direction.

For $D_2$ the free eigenvalues take the form  (for $\rho = 1$)
\begin{equation}
\label{D2_free_spectrum_1}
\lambda_1 = \pm \sqrt{ A_1 - p^2 \pm  2 i \sqrt{A_1 p^2} },
\end{equation}
and
\begin{equation}
\label{D2_free_spectrum_2}
\lambda_2 = \pm \sqrt{ A_2 - p^2 \pm  2 i \sqrt{A_2 p^2} }
\end{equation}
in which the $\pm$ signs are chosen independently and the eigenvalues are doubly degenerate, and having 
defined
\begin{equation}
p^2 = \sum_{\mu=1}^4 \sin^2 p_\mu,
\end{equation}
\begin{equation}
A_1 = \frac{ c_1^2 c_2^2 +  c_1^2 c_3^2+  c_1^2 c_4^2 +  
c_2^2 c_3^2 +  c_2^2 c_4^2 +  c_3^2 c_4^2}{3} - 2 c 
= 0 + \mathcal{O}(a^2),
\end{equation}
\begin{equation}
A_2 = \frac{ c_1^2 c_2^2 +  c_1^2 c_3^2+  c_1^2 c_4^2 +  
c_2^2 c_3^2 +  c_2^2 c_4^2 +  c_3^2 c_4^2}{3} + 2 c  
= 4 + \mathcal{O}(a^4),
\end{equation}
where $c_\mu = \cos p_\mu$, and $c = c_1 c_2 c_3 c_4$. 
Expanding for small momenta gives
\begin{equation}
\lambda_1 = \pm \sqrt{ - p^2 }  =  \pm i p, \qquad
\lambda_2 = \pm 2 \sqrt{ 1 \pm i p }  =  \pm 2 \pm i p,
\end{equation}
so that at low momenta, the eigenmodes corresponding to 
$\lambda_2$ get a mass of $\pm 2$, while the eigenmodes 
corresponding to $\lambda_1$ are massless.

Finally, the free spectrum of $D_4$ reads:
\begin{equation}
\label{D4_free_spectrum}
\lambda_1 =  - c \frac{\rho}{a} \pm i \sqrt{p^2}, \quad
\lambda_2 =  + c \frac{\rho}{a} \pm i \sqrt{p^2}, \quad
\end{equation}
Note that, in the continuum limit, the point where the spectrum 
of the $\dks$ operator intersects the real axis corresponds to 
four massless modes. By contrast, $D_2$ leads to one mode in each 
of the two intersections away from the origin, and two at the 
origin. Finally, for $D_4$ one obtains two modes at each of the 
two intersections of the spectrum with the real axis.

\begin{figure}
\centerline{
\includegraphics[width=0.50\textwidth]{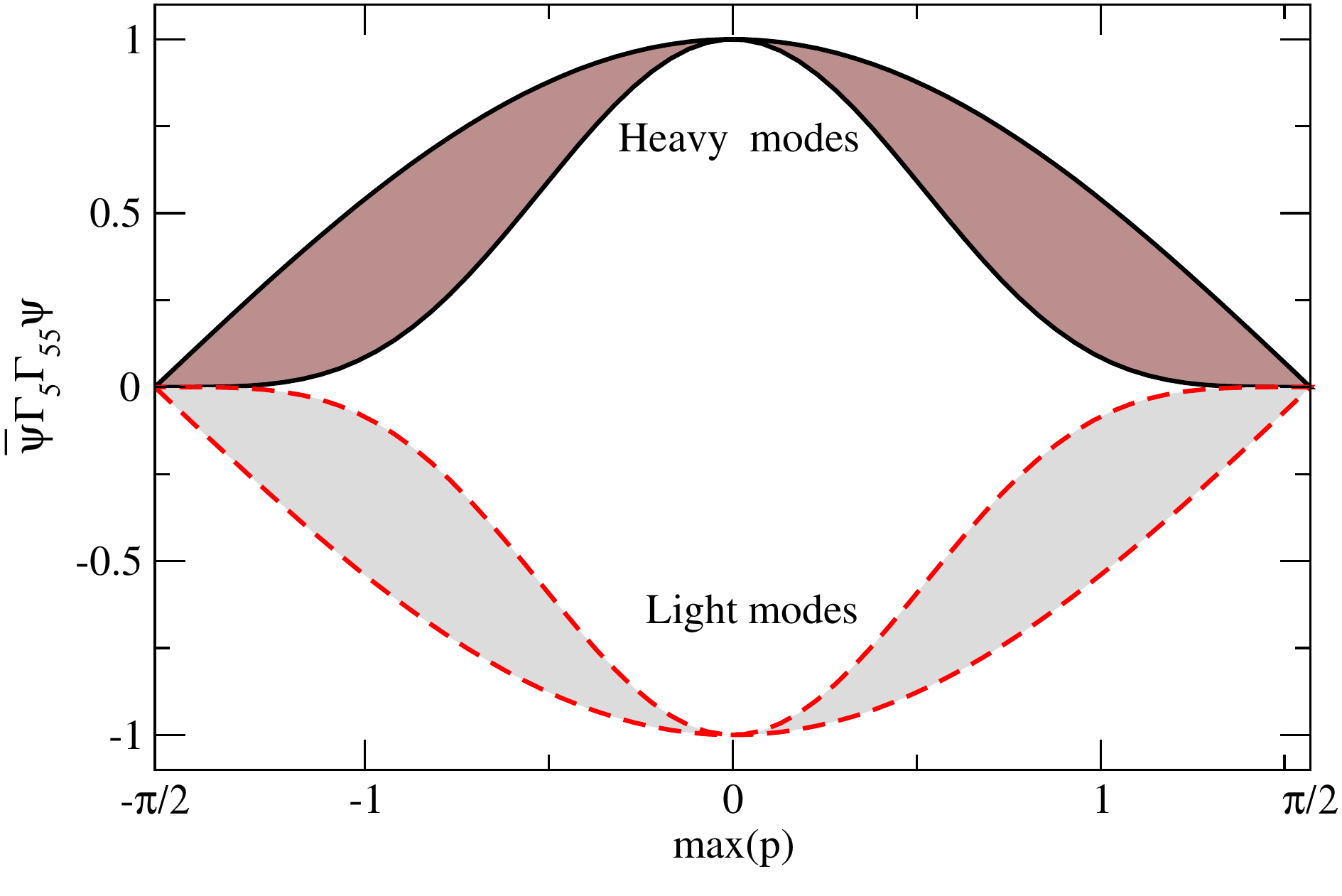}
\includegraphics[width=0.50\textwidth]{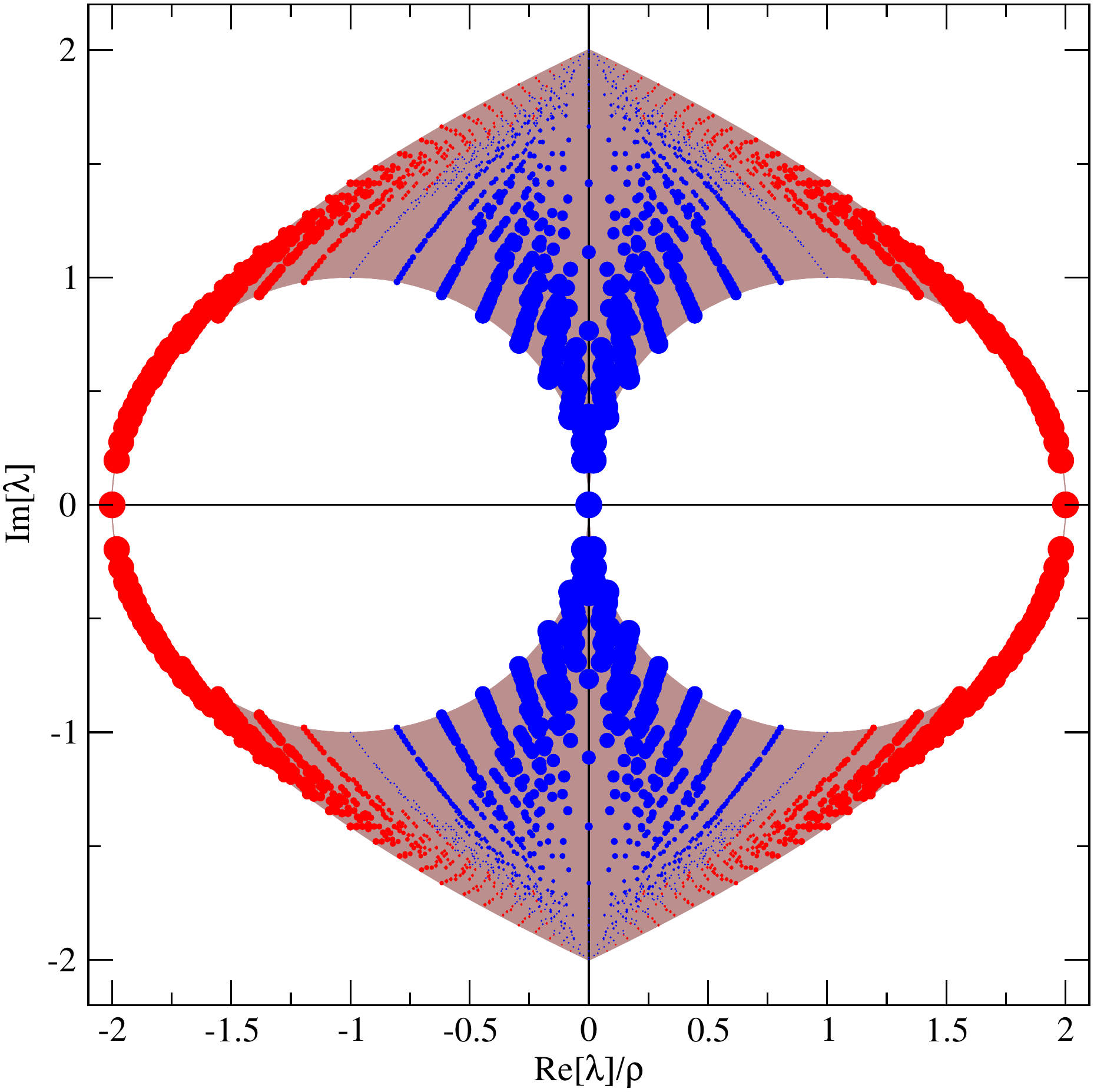}
}
\label{lip}
\caption{(Left) Taste chirality properties of the $D_2$ and $D_4$ eigenvectors, as a 
function of (the component of minimum modulus of) their momentum, 
in the free limit. On an infinite lattice, the eigenvectors 
associated with real eigenvalues have vanishing momentum and a 
well-defined taste chirality $\pm 1$. For eigenmodes corresponding to 
eigenvalue $\lambda_1$, the taste chirality becomes $+1$, while for 
$\lambda_2$-eigenmodes the taste chirality approaches $-1$. (Right) 
The taste chiralities of the eigenmodes of the $D_2$ operator; the size
of the points corresponds to the magnitude of $c$, while the color 
indicates the sign: blue for $+c$, red for $-c$.}
\end{figure}

The taste chirality of the eigenmodes $\Psi$ of $D_4$ and $D_2$, is given 
by  $(\bar\Psi \Gamma_{55} \Gamma_5 \Psi)$, 
where $\Gamma_{55}=\gamma_5\otimes\gamma_5$ exactly, and 
$\Gamma_5 = \gamma_5\otimes {\bf 1} + {\cal O}(a)$ in spin $\otimes$ taste.
The taste chirality of the eigenmodes corresponding to eigenvalues 
$\lambda_1$ is $c$, while that of the $\lambda_2$-eigenvectors is $-c$.
This is also depicted in Fig.~\ref{lip};
one can see that the taste chirality of the real eigenmodes is $\pm 1$, and is the same ($+1$ or $-1$)
for all modes in a given branch of the $D_4$ or $D_2$ spectrum. The implications of a well-defined
taste chirality have been stressed in \cite{staggeredoverlap2}: if $\Gamma_{55} \Gamma_5 \approx \pm 1$,
then $\Gamma_{55} \approx \pm \Gamma_5$, so that the {\em spin} chirality of the real eigenmodes can
be probed by $\Gamma_{55}$. This is the reason why the index theorem applies to $D_2$ and $D_4$, while
it does not for $\dks$ (where both the $\pm 1$ taste chiralities lay on the same single branch.)

A shift of the spectra by a real value can thus lead to chiral 
low-momentum zero modes in each branch, and hence to the possibility 
of constructing an appropriate index. A common way to 
study the index consists of looking at the flow of eigenvalues 
$\lambda(m)$ of:
\begin{equation}
H(m) = \gamma_5 (D + m).
\label{standardflow}
\end{equation}
In general, if $(D + m)$ has a zero-mode $|\Psi_0\rangle$ for 
$m=m_0$, then, correspondingly, $H$ has a vanishing eigenvalue 
$\lambda(m_0)=0$. With a small perturbation of $m$ away from $m_0$, 
i.e. $m=m_0+\delta m$, at leading order the eigenvalues get 
displaced by an amount  $\langle\Psi_0| \gamma_5 (m-m_0) |
\Psi_0\rangle$, namely one finds \emph{crossings}  $\lambda(m)= 
\pm (m-m_0)$, if $|\Psi_0\rangle$ is a chiral mode: 
$\langle\Psi_0| \gamma_5 |\Psi_0\rangle = \pm 1$. 
As pointed out in \cite{Adams:2011xf},
the saturation of $(\bar\Psi \Gamma_{55} \Gamma_5 \Psi)$ at value $\pm 1$
discussed above allows us to trade $\Gamma_5$ for $\Gamma_{55}$ and use the latter in
eq.(\ref{standardflow}).

An alternative 
way to look at the spectral flow was proposed in 
ref.~\cite{staggeredoverlap1} for the $D_4$ operator, by studying 
the eigenvalues of\footnote{Actually, Ref.~\cite{staggeredoverlap1} proposed to consider the
spectral flow of $(i \dks - \frac{\rho}{a} \Gamma_5)$. As recognized in \cite{staggeredoverlap2},
that operator is the same as eq.(\ref{Adamsflow}) up to a redefinition of the $\eta_\mu$ phase factors.}
\begin{equation}
\hat{H}(\rho) = \Gamma_{55} \dks - \frac{\rho}{a} \Gamma_{5} .
\label{Adamsflow}
\end{equation}
Fig.~3 displays a comparison of the 
two different ways to define the spectral flow for the $D_4$ 
operator (see \cite{Follana:2011kh} for a recent related study): the plots in the top row show the flow of eigenvalues 
of $\hat{H}$ as a function of $\rho$ (eq.(\ref{Adamsflow})), 
whereas those in the bottom row refer to the ``standard'' 
definition of the flow, using eq.~(\ref{standardflow}). In each 
row, the left panel displays the results from a cold (i.e., free) 
configuration on a lattice of size $16^3 \times 32$, while the 
central panel is obtained from a cooled configuration of 
topological charge $Q=1$ on a lattice of size $8^4$, and finally 
the right panel displays the results from a ``rough'' (i.e., 
non-cooled) quenched $Q=-1$ configuration at $\beta=6/g^2=6$, on a 
lattice of size $12^4$. In the latter case, the comparison of the 
two flow definitions shows that, with the standard definition, 
the region around the real axis is populated by a large number of 
eigenvalues, preventing one from identifying the crossing with 
accuracy.

\begin{figure}
\centerline{
\includegraphics[width=0.39\textwidth]{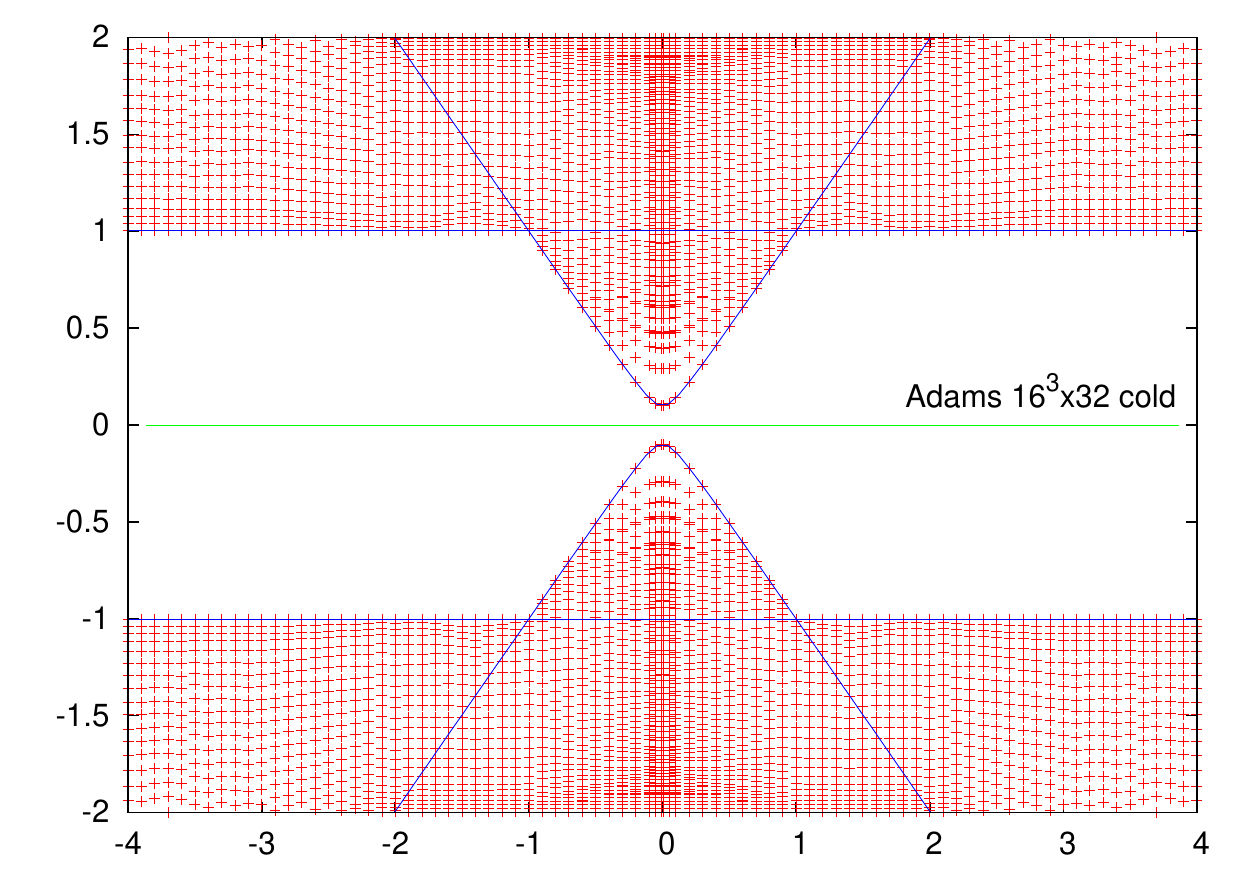} \hfill 
\includegraphics[width=0.39\textwidth]{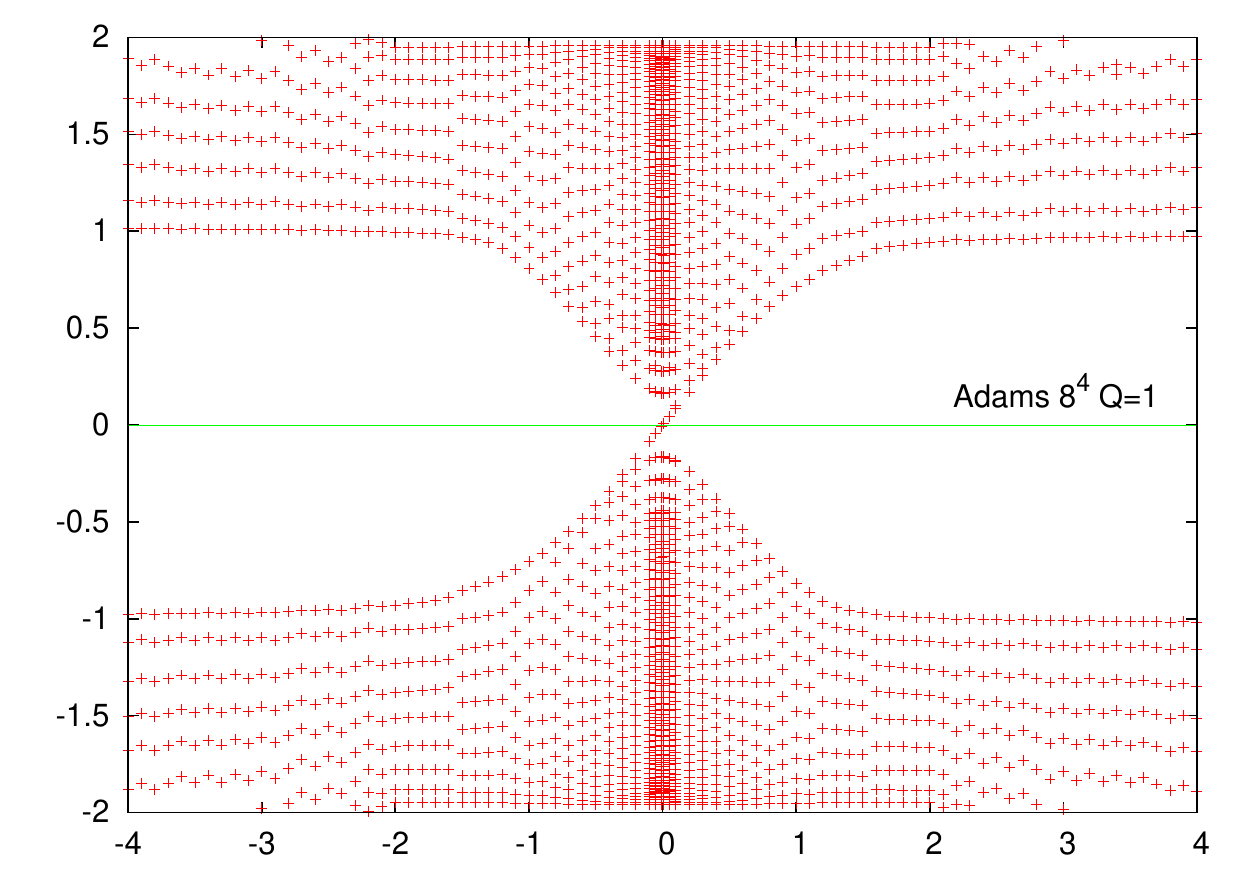} \hfill 
\includegraphics[width=0.39\textwidth]{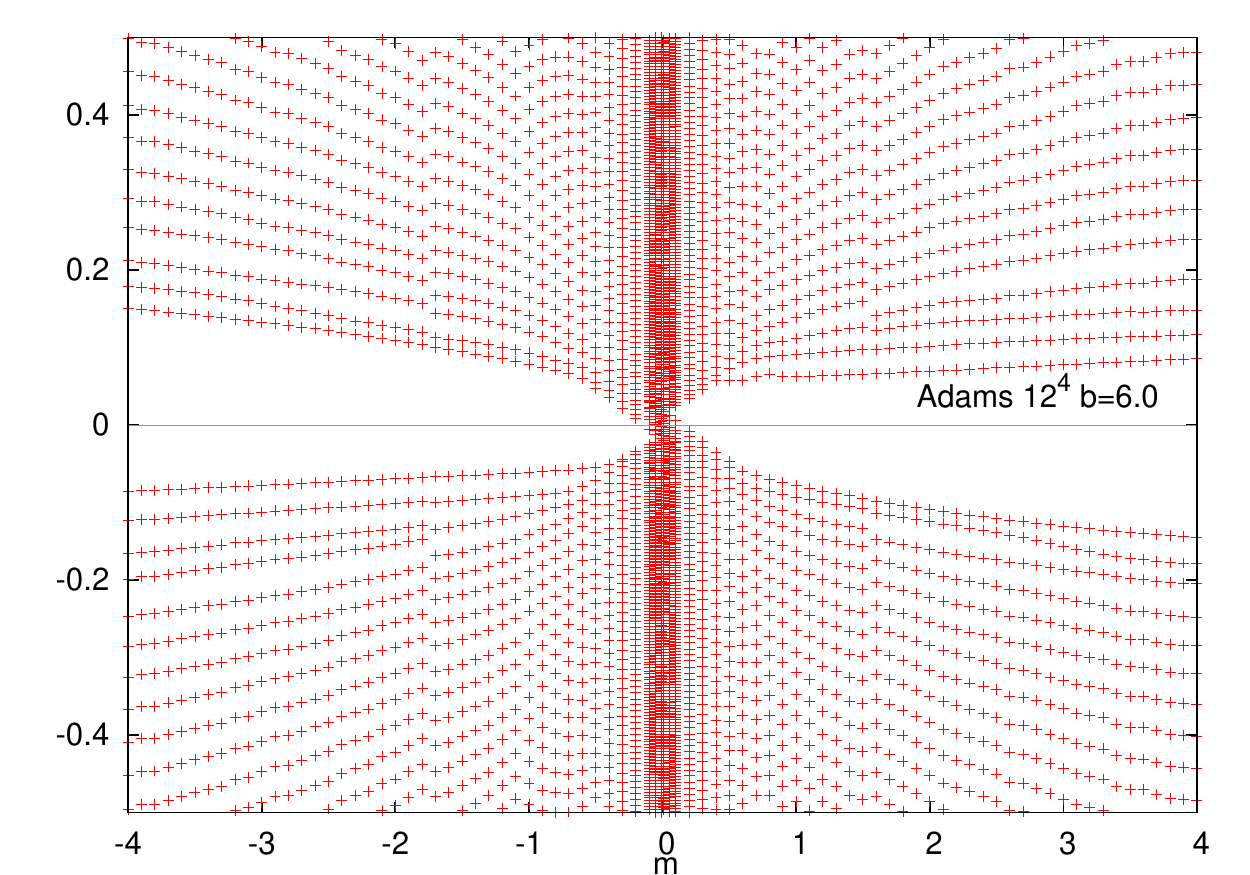}
}
\centerline{
\includegraphics[width=0.39\textwidth]{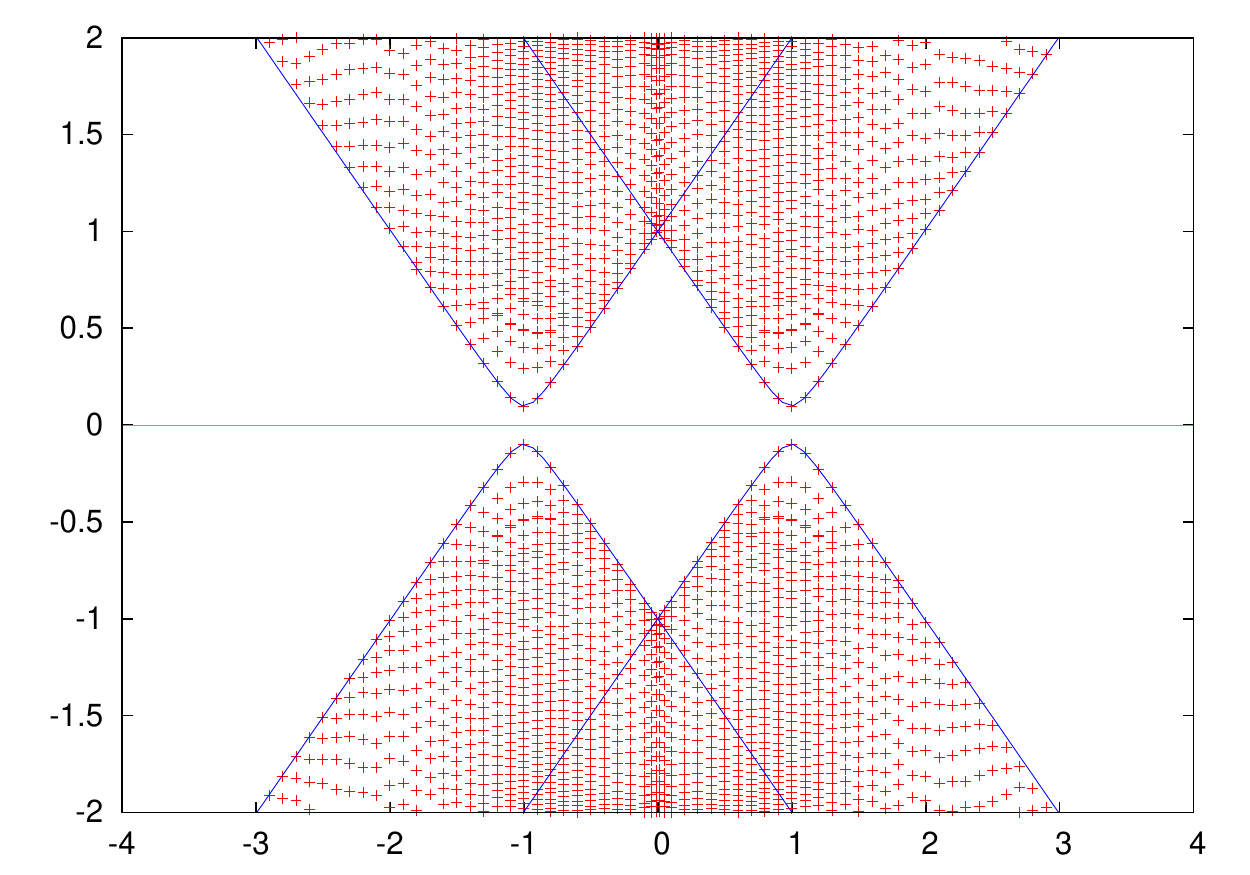} \hfill 
\includegraphics[width=0.39\textwidth]{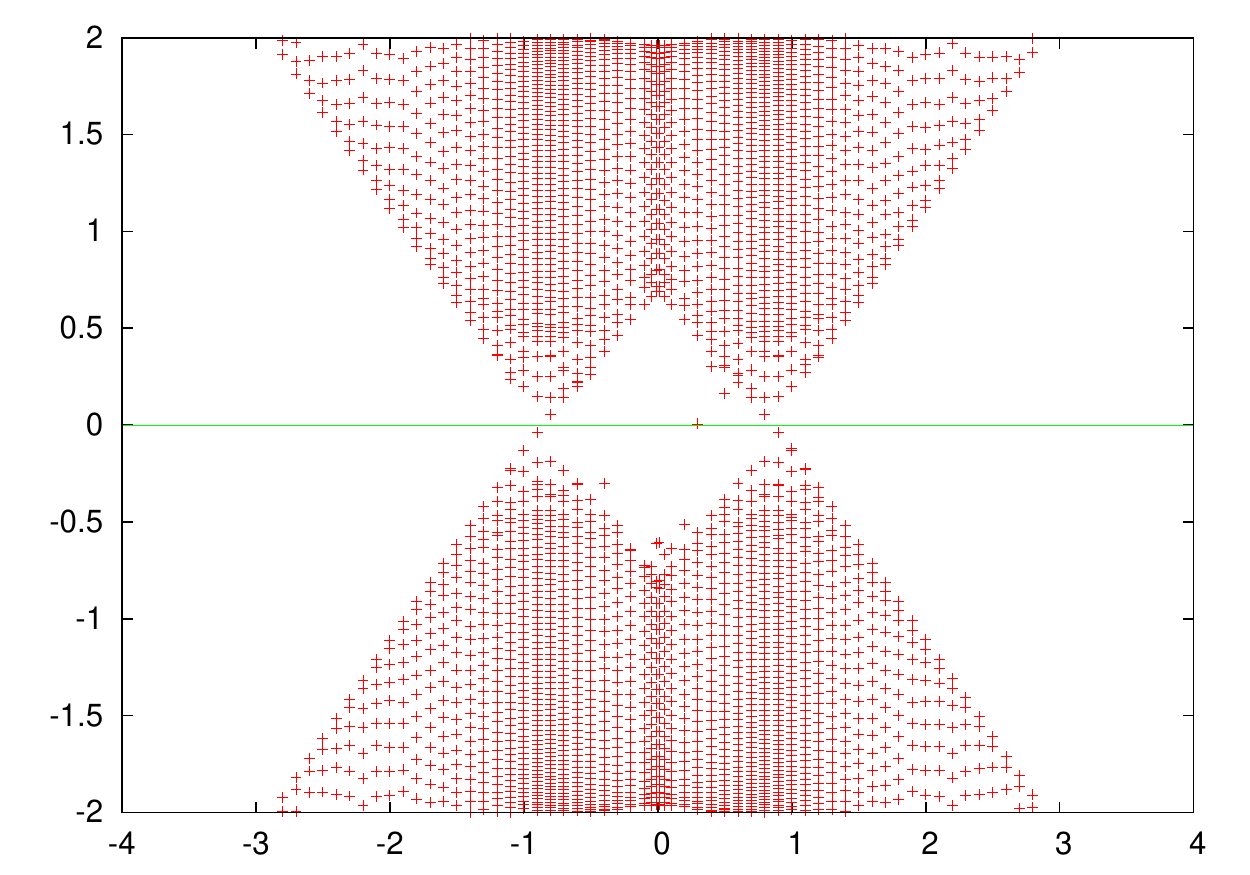} \hfill 
\includegraphics[width=0.39\textwidth]{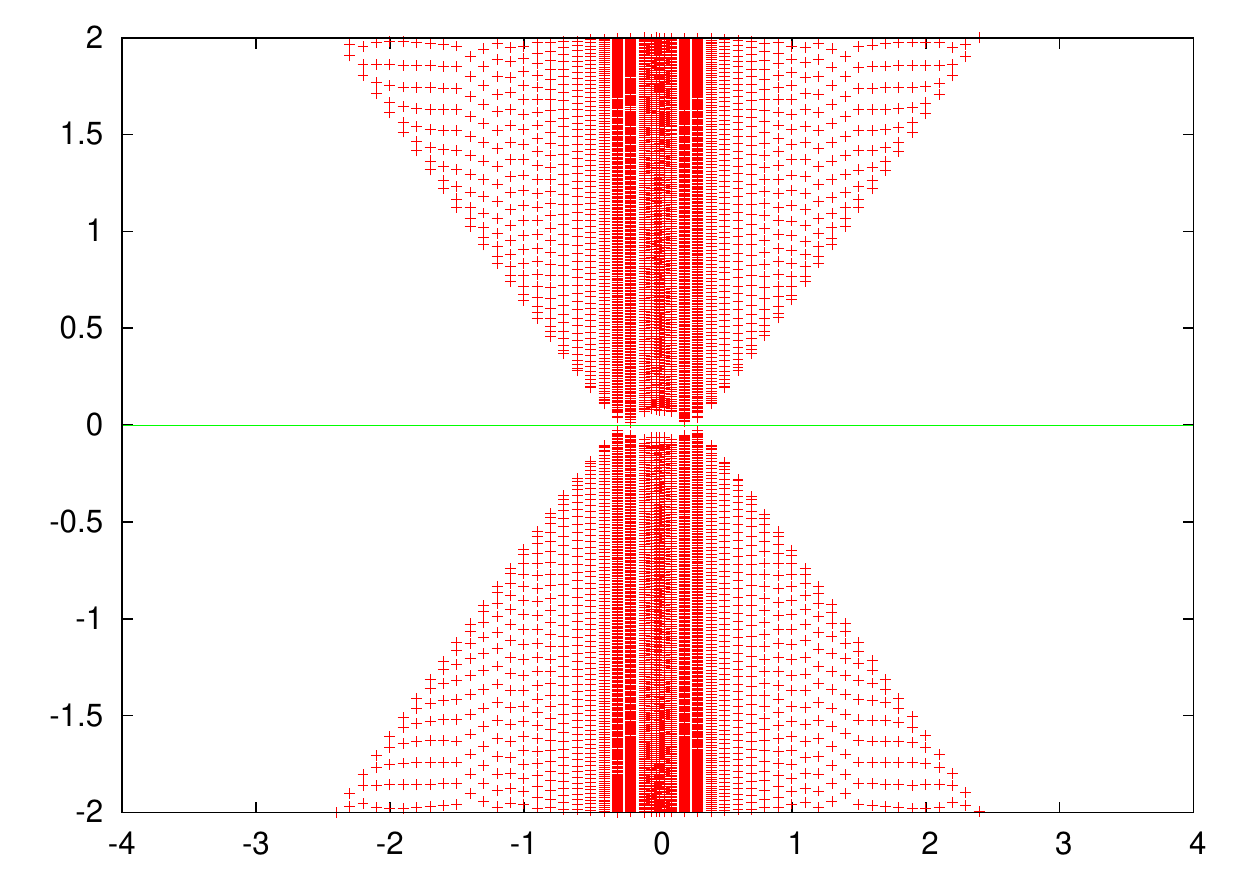}
}
\label{Adams_original_classic}
\caption{Comparison of the spectral flow for the $D_4$ operator 
as obtained from the eigenvalues of the operators defined in 
eq.~(\protect\ref{Adamsflow}) (top row panels) and 
eq.~(\protect\ref{standardflow}) (bottom panels). The three plots 
(from left to right) in each row show, respectively, the 
eigenvalues of $\tilde{H}$ (or $H$) as a function of $\rho/a$ (or 
$m$) from a free configuration on a lattice of size $16^3 \times 
32$, from a cooled configuration of topological charge $Q=1$ on a 
lattice of size $8^4$, and from a non-cooled $Q=-1$ gauge 
configuration at $\beta=6$, on a lattice of size $12^4$.}
\end{figure}

\begin{figure}
\centerline{\includegraphics[width=0.39\textwidth]{flow_cf1632_cold_Adams_classic.pdf} \hfill 
\includegraphics[width=0.39\textwidth]{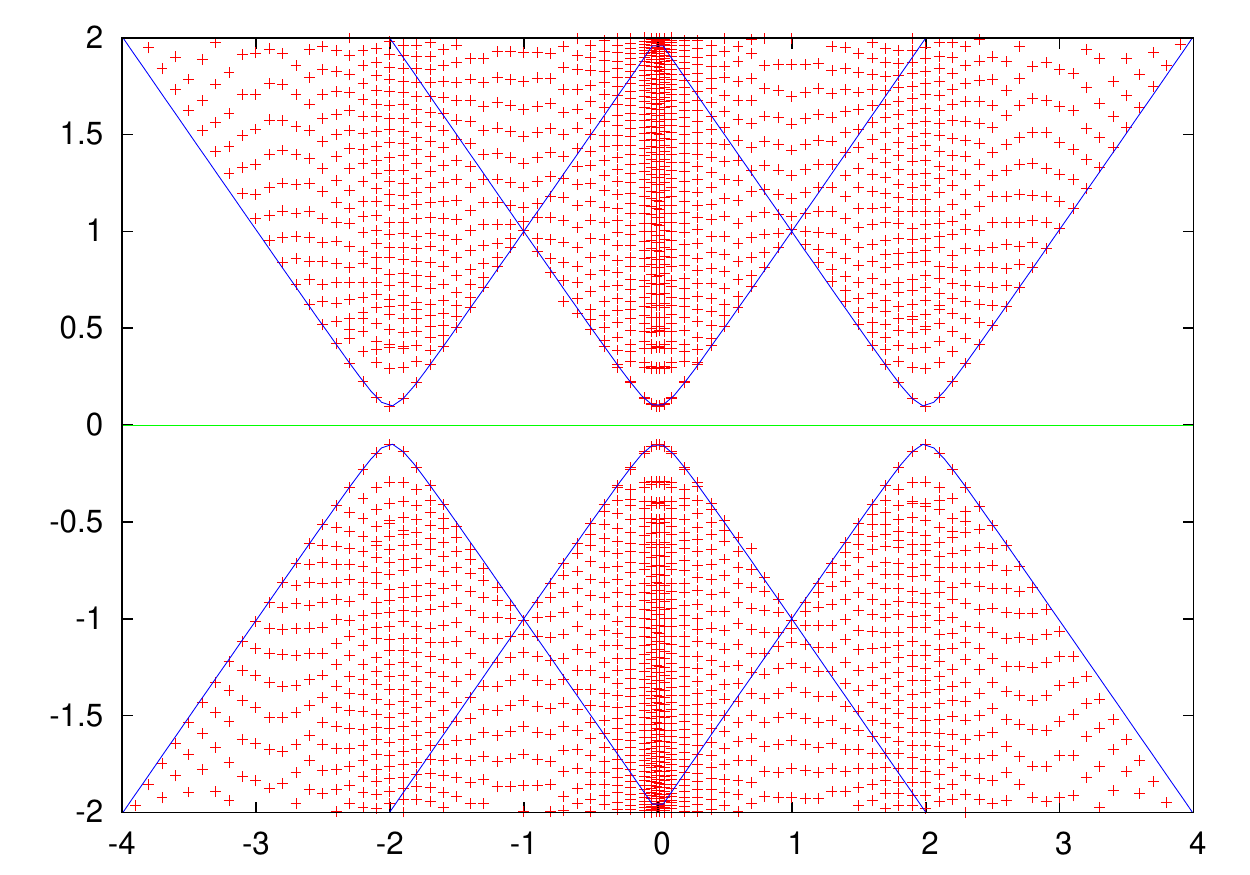} \hfill 
\includegraphics[width=0.39\textwidth]{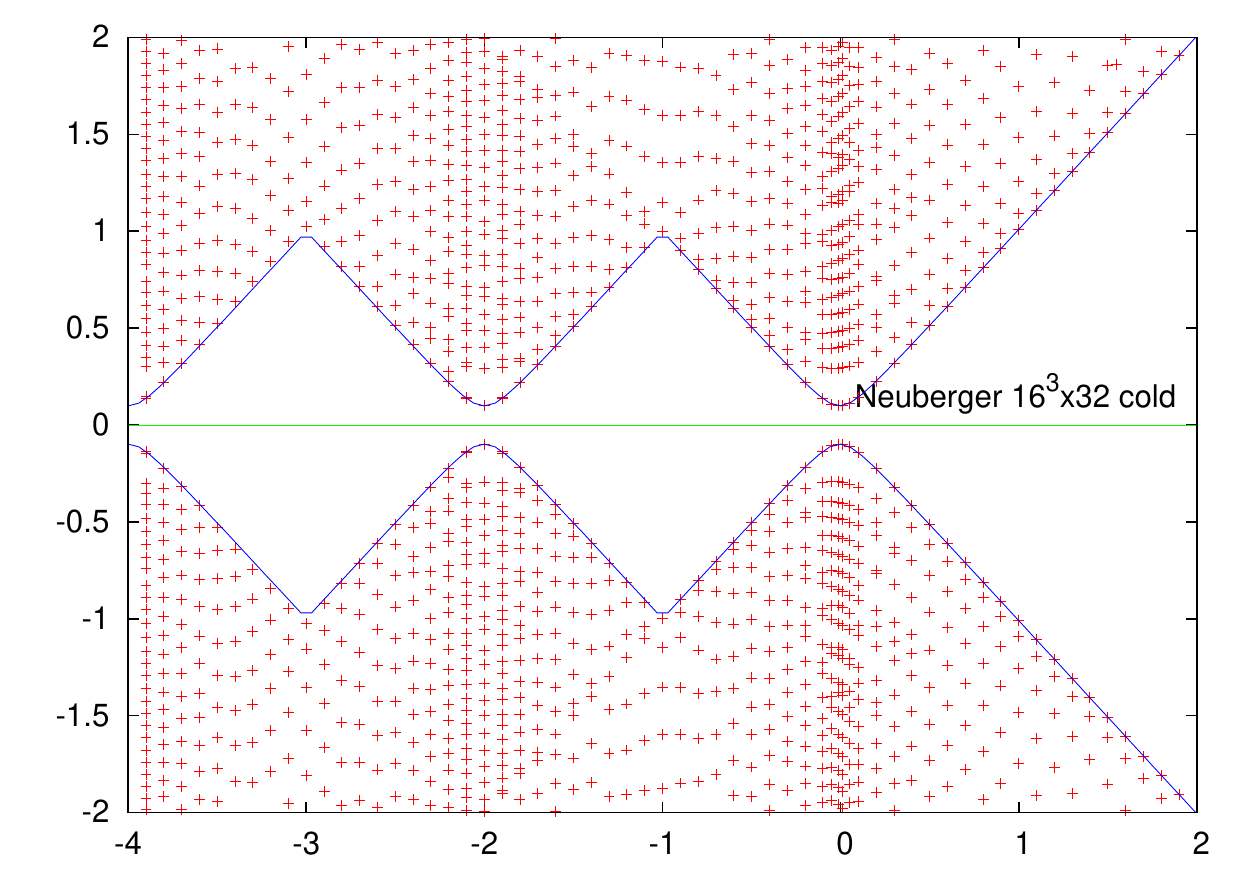}
}
\centerline{
\includegraphics[width=0.39\textwidth]{flow_cf_84_Q1_Adams_classic.pdf} \hfill 
\includegraphics[width=0.39\textwidth]{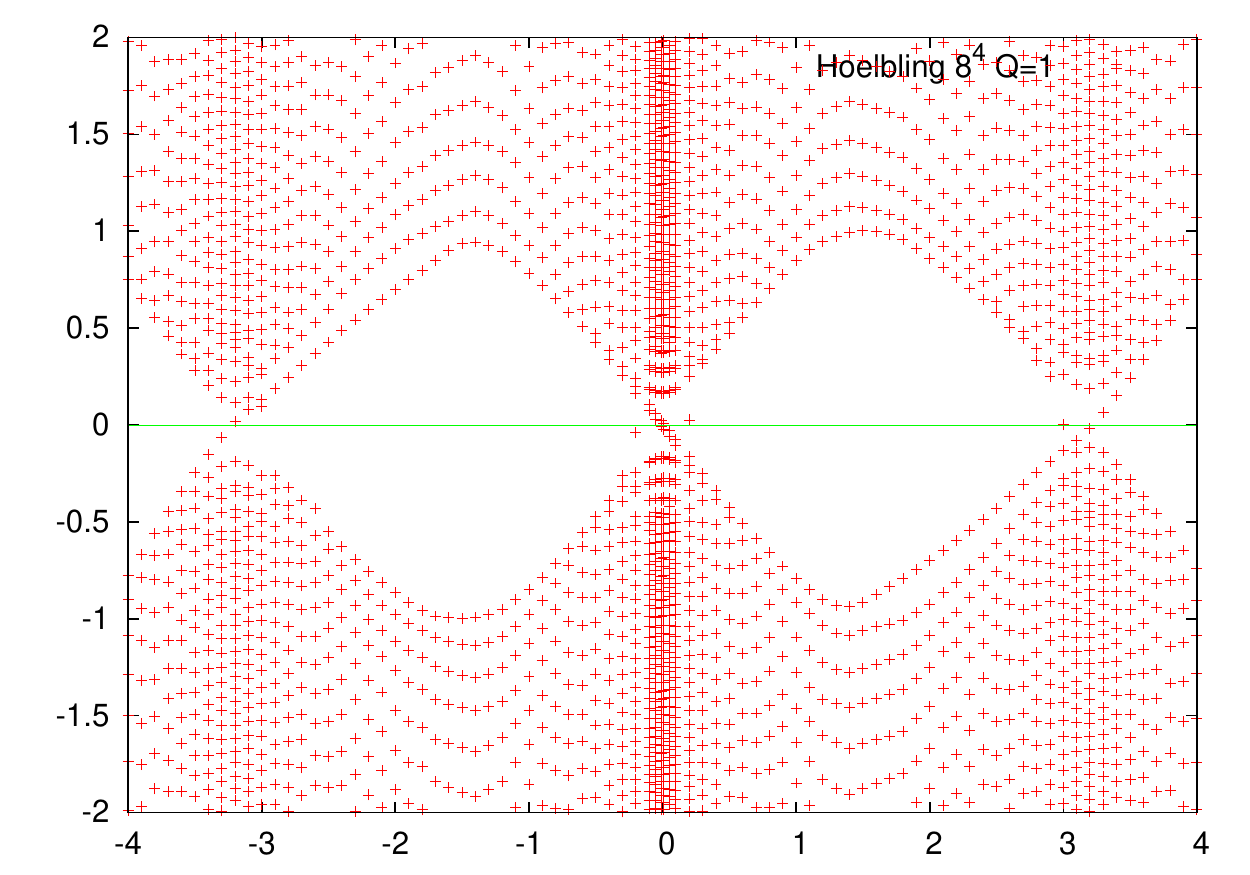} \hfill 
\includegraphics[width=0.39\textwidth]{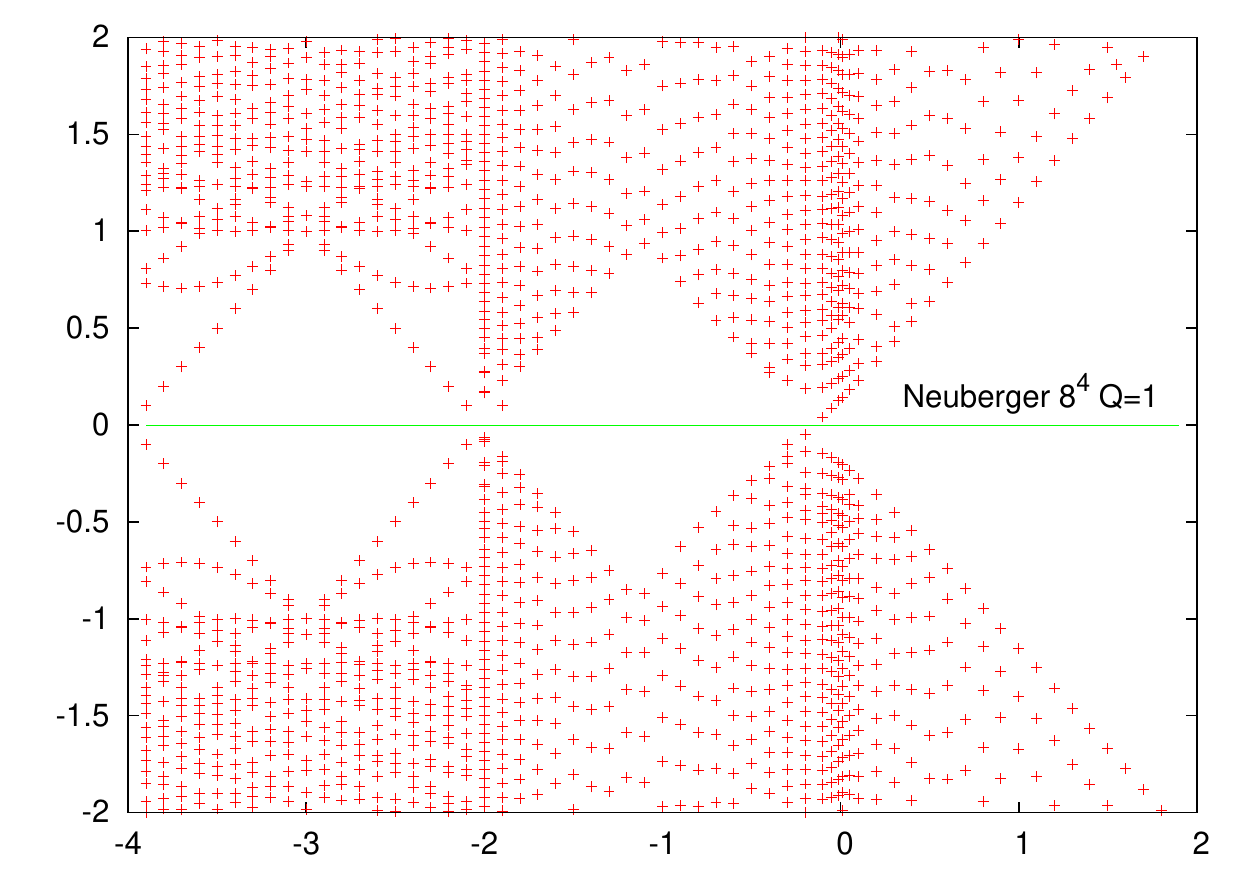}
}
\centerline{
\includegraphics[width=0.39\textwidth]{flow_cf124_b60_Adams_classic.pdf} \hfill 
\includegraphics[width=0.39\textwidth]{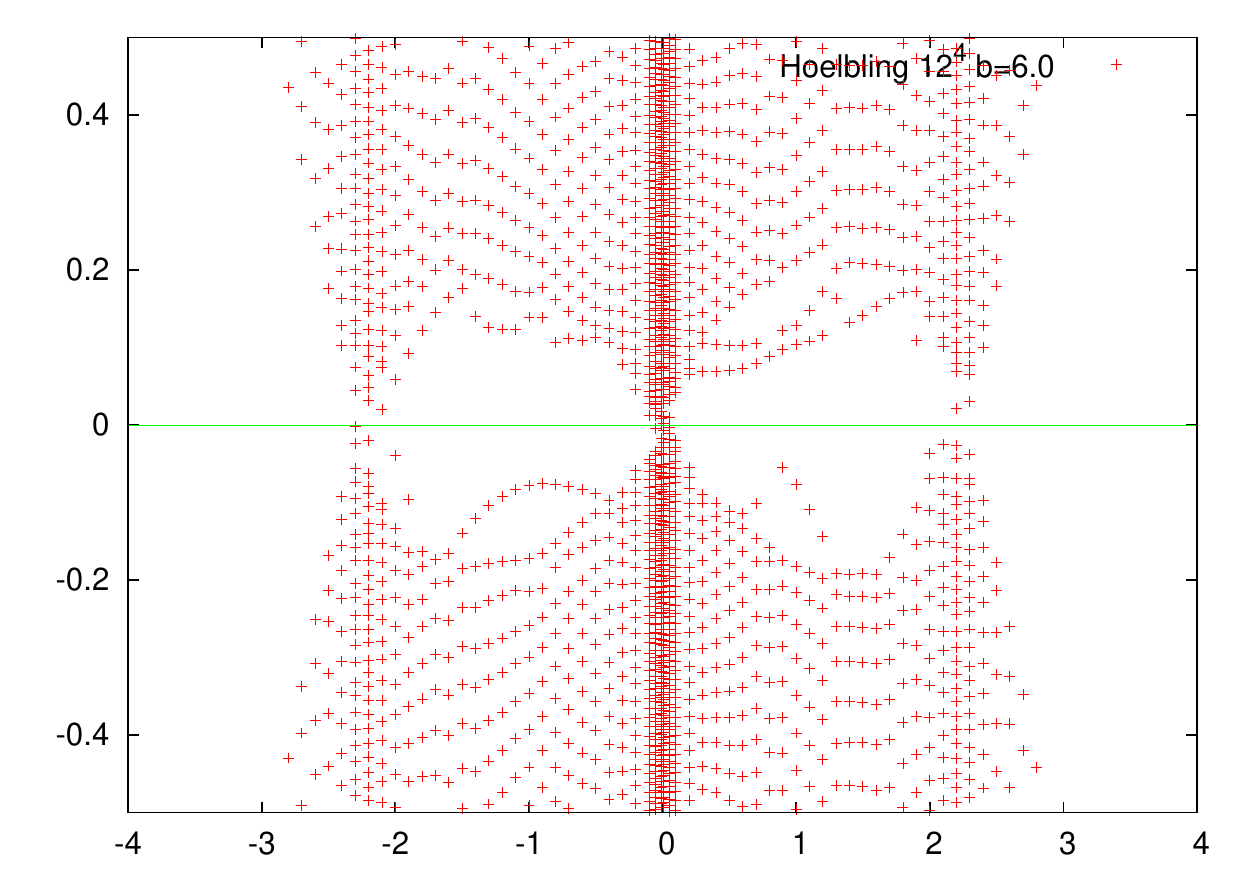} \hfill 
\includegraphics[width=0.39\textwidth]{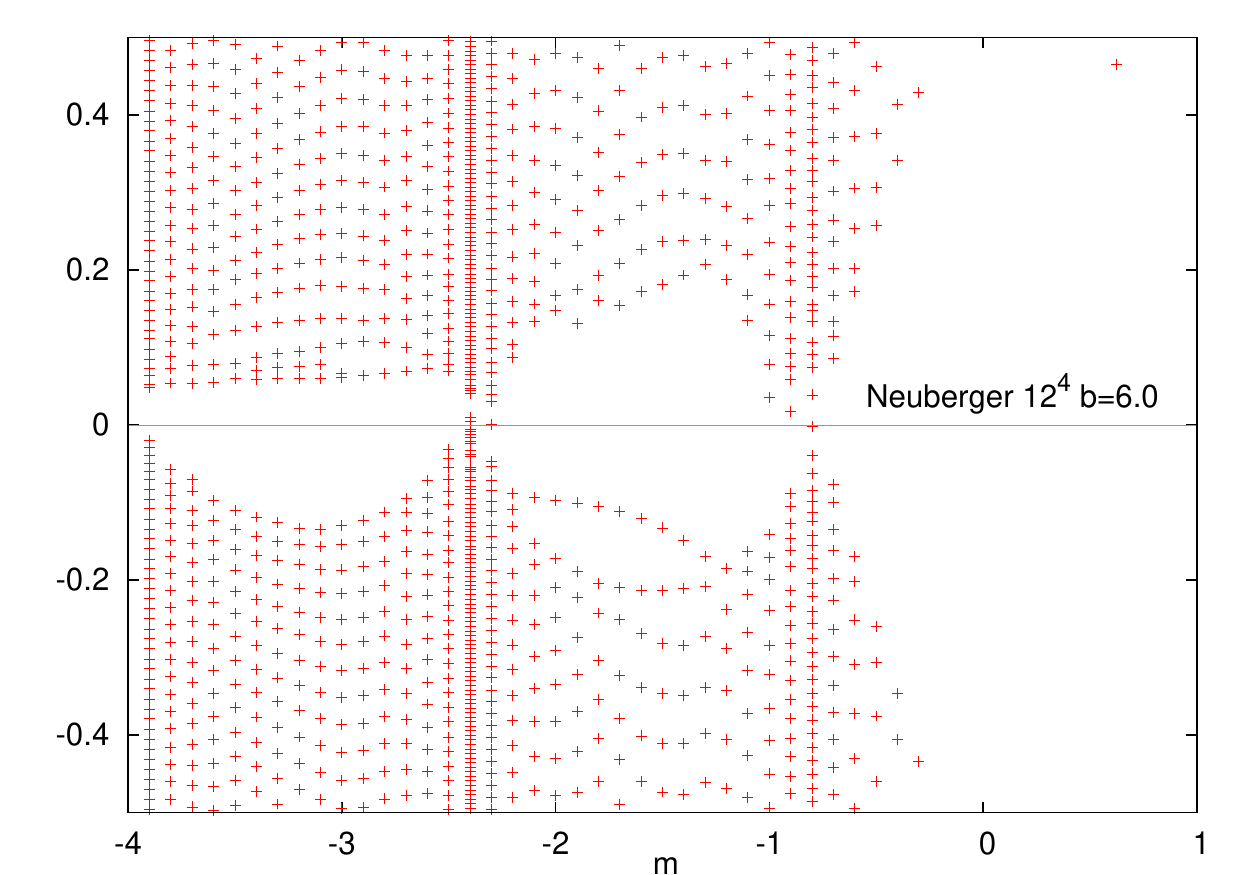}
}
\caption{Spectral flows for $D_4$ (left), $D_2$ (center) and a 
standard Wilson operator (right), on a cold configuration (top), 
on a cooled $Q=1$ configuration (middle), and on a non-cooled 
$Q=-1$ quenched configuration at $\beta=6$ (bottom).
The solid blue lines in the top row show analytic results.}
\label{flowcomparison}
\end{figure}

Next, it is interesting to compare the identification of the 
index, using the spectral flow defined from 
eq.~(\ref{standardflow}), for staggered fermions with a taste-dependent mass term, and for conventional Wilson fermions. This 
is shown in Fig.~\ref{flowcomparison}: the left, central and 
right plot in each row show the spectral flow for $D_4$, $D_2$ 
and a standard Wilson operator, respectively, while the three 
different rows, from top to bottom, refer to a cold 
configuration, to a cooled $Q=1$ configuration, and to a non-
cooled $Q=-1$ quenched configuration at $\beta=6$. It is 
interesting to observe that, as expected, the spectral flow on a 
cooled instanton configuration clearly reveals $N_f \times Q$ 
crossings. However, one already sees that in the plots of the 
$\beta=6$ configuration the gap tends to close. This is 
especially the case for the $D_4$ operator, and is related to the 
properties that will be discussed in Section~\ref{sec:numerical}.

The overall message that can already be drawn from these 
observations (before addressing a full-fledged numerical 
investigation) is that the gauge field fluctuations in 
interacting configurations reduce the width of the gap in the 
spectrum, and blur the distinction between light modes and 
doublers.

\section{Numerical investigation on interacting configurations}
\label{sec:numerical}
\begin{figure*}
\centerline{\includegraphics[width=0.275\textwidth,clip=true]{StaggeredBatman.pdf} \hfill 
\includegraphics[width=0.265\textwidth]{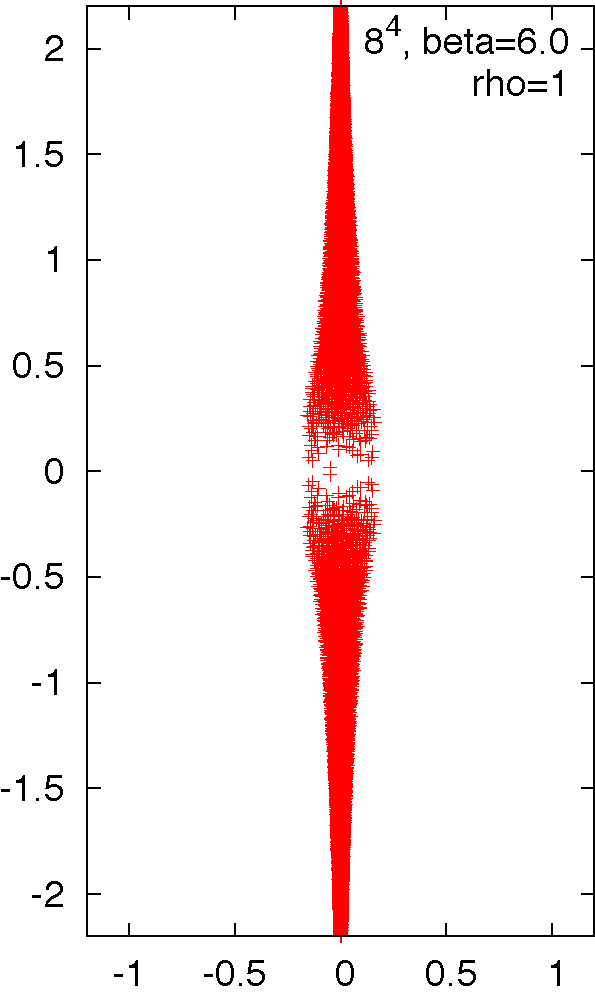} \hfill 
\includegraphics[width=0.265\textwidth,clip=true]{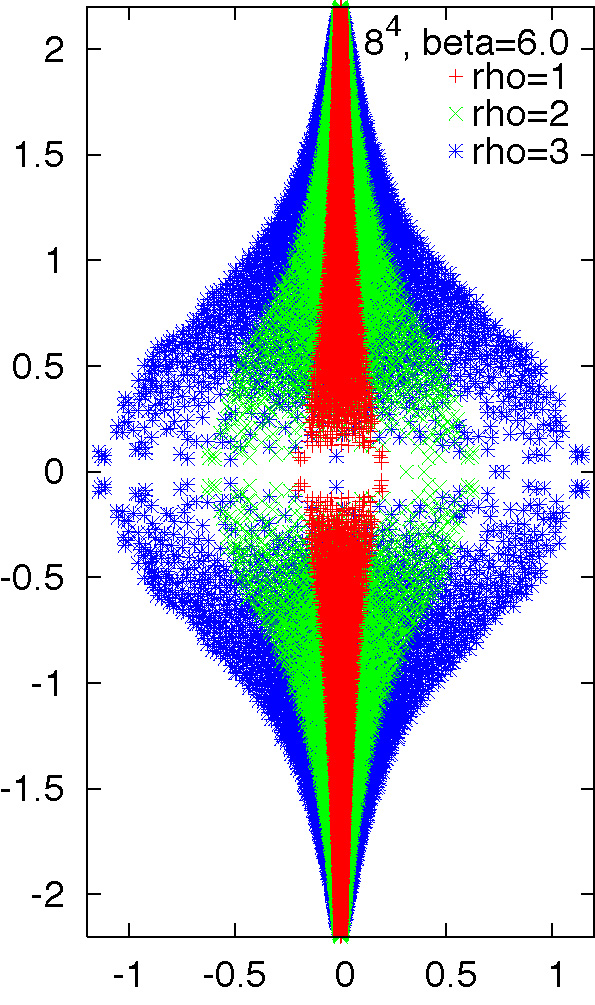} \hfill 
\includegraphics[width=0.265\textwidth,clip=true]{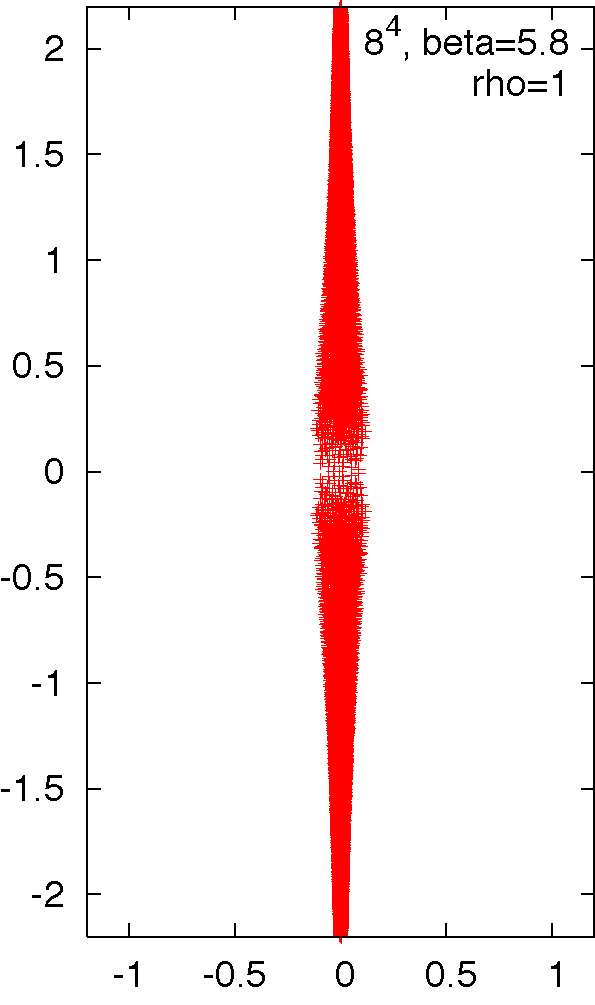}
}

\centerline{
\includegraphics[width=0.235\textwidth,viewport=-1 5 495 402]{HoelblingBatman.pdf}\hspace{0.3cm}
\includegraphics[width=0.248\textwidth]{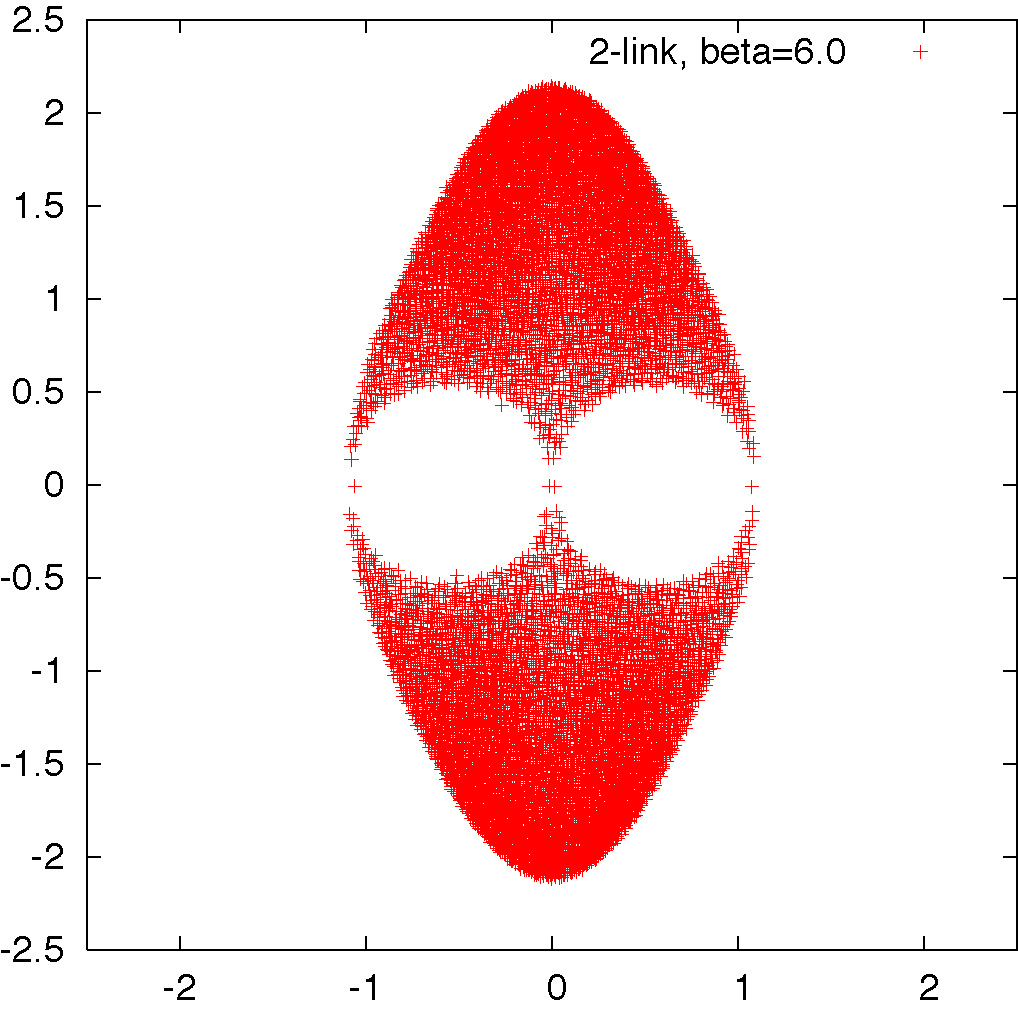}\hspace{0.3cm}
\includegraphics[width=0.248\textwidth,clip=true]{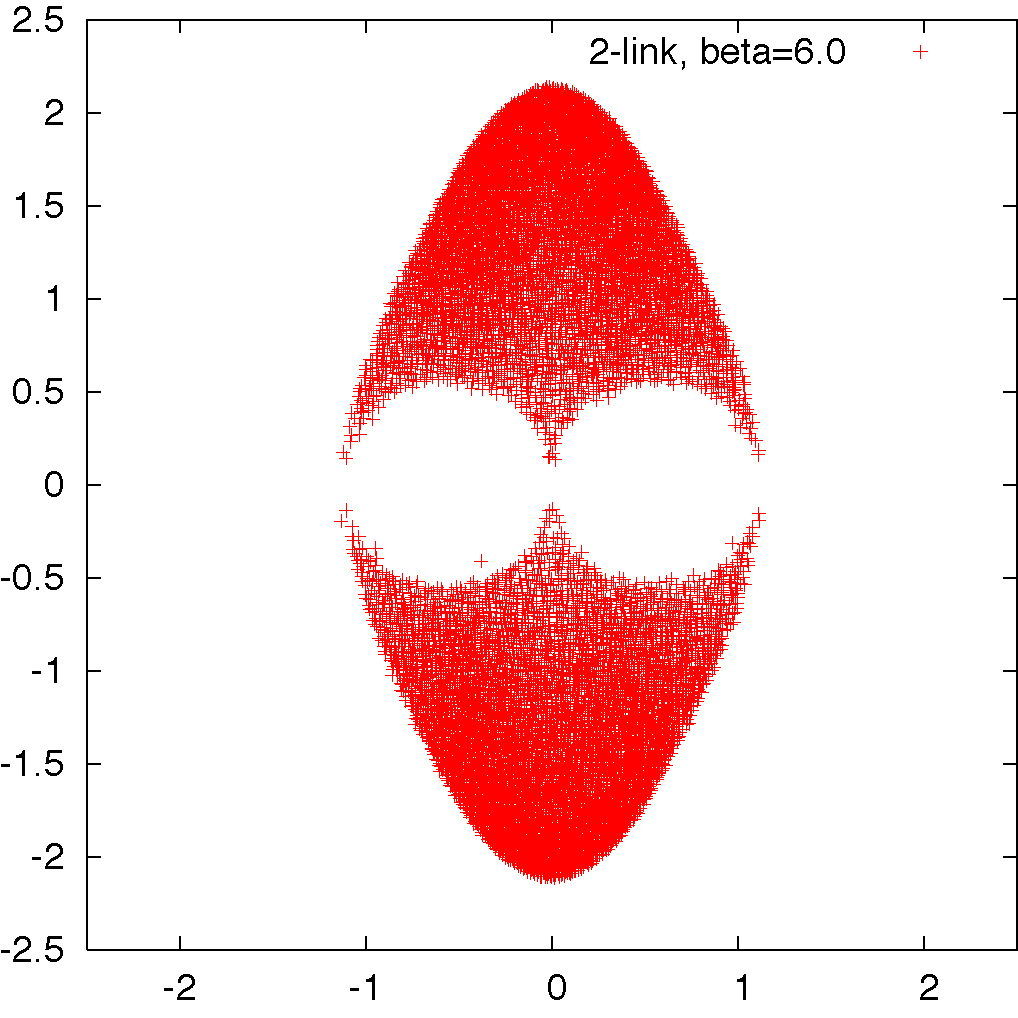}\hspace{0.3cm}
\includegraphics[width=0.248\textwidth,clip=true]{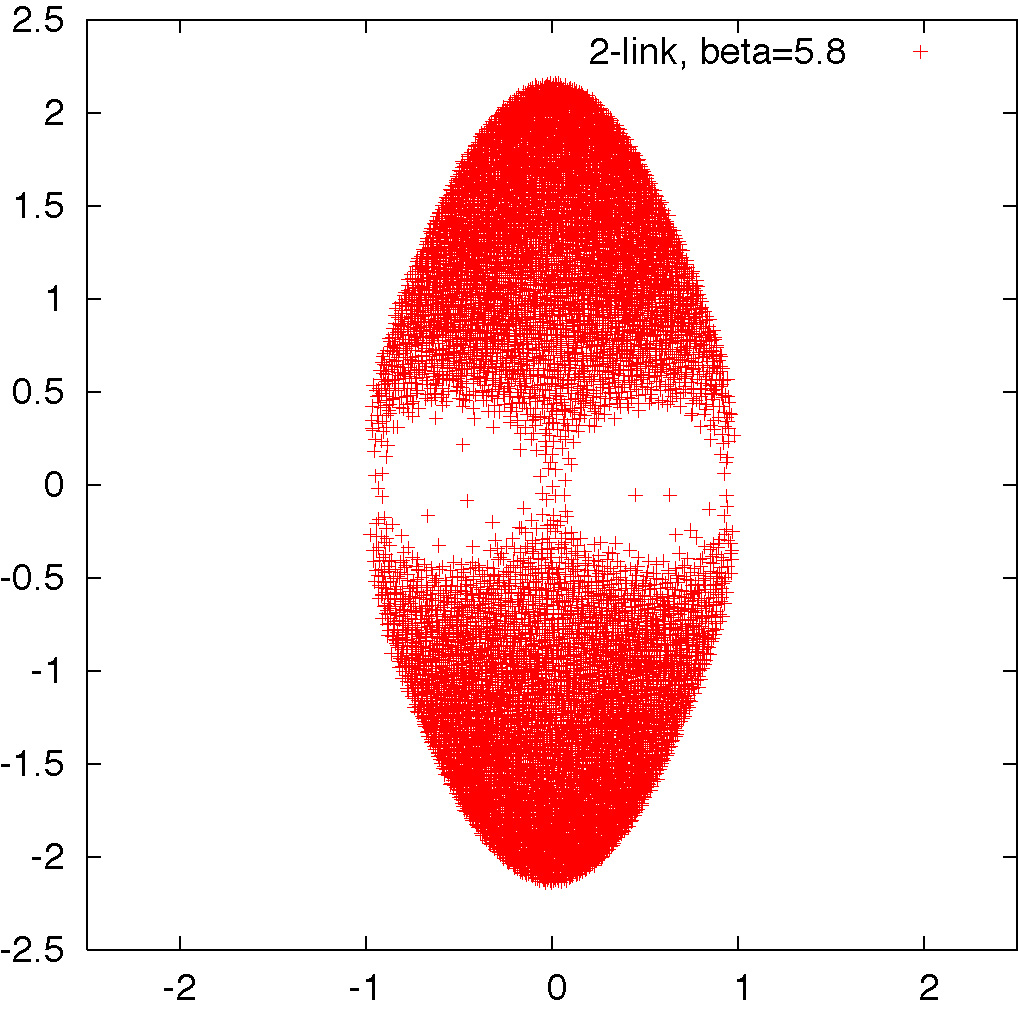}
}
\caption{Spectra of $D_4$ (top) and $D_2$ (bottom) on different 
types of configurations. As compared to the free case (left), the 
gap in the spectrum of eigenvalues of $D_4$ on interacting 
configurations tends to close more rapidly than in the case of 
$D_2$. The second and third plot in each row are obtained from 
quenched configurations at $\beta=6$ (in the third plot on the 
top row, symbols of different colors correspond to different 
values of $\rho/a$). Finally, the plots on the right are obtained 
from a coarser lattice, at $\beta=5.8$ (roundoff errors cause some
breaking of the complex conjugation symmetry of the spectrum).}
\label{Adams_vs_Hoelbling}
\end{figure*}

As we showed in the previous section, the fluctuations in typical 
interacting configurations lead to a filling of the gap in the 
spectral flow for the various lattice Dirac operators that we are 
considering, making a proper identification of the index 
difficult. A related effect can also be seen directly in the 
spectra of the operators: the panels in
Fig.~\ref{Adams_vs_Hoelbling} show a comparison of the spectra of 
$D_4$ (top row) and $D_2$ (bottom row) in the free case (left), 
and in interacting configurations at $\beta=6$ (central panels, in which different values of $\rho/a$ or $m$ are used) and at 
$\beta=5.8$ (right). The figure shows evidence for the superior 
robustness of lattice fermions based on the $D_2$ operator, over 
$D_4$: at $\beta=5.8$ for example, a gap remains clearly visible 
for $D_2$, while it has all but disappeared for $D_4$.

This can be understood from the fact that, since $D_4$ involves 
4-link parallel transporters, it is more sensitive to the gauge 
field fluctuations in interacting configurations than $D_2$ which 
involves 2-link transporters only\footnote{Note that the same reason also 
explains the fact that the chirality of near-zero modes of the ordinary 
staggered operator is typically small~\cite{deForcrand:1998ng}.}.

However, for practical applications in large-scale simulations, 
it is important to remark that, as usual, the effect of gauge 
fluctuations can be considerably reduced through some suitably 
optimized smearing procedure.

Next, we considered the effectiveness of these operators for 
spectroscopy calculations. To this end, we performed a simple 
test, by studying the mass $m_{PS}$ of the lightest meson in the 
pseudoscalar channel (the pion). We computed the quark propagator 
$G(x,y,z,t)$ from a point source, on quenched configurations at 
$\beta=6$ on a lattice of size $16^3 \times 32$, then we 
evaluated the $\vec{p}=\vec{0}$ correlation function
\begin{equation}
C(t) = \sum_{xyz} G(x,y,z,t) \Gamma_{55} G(x,y,z,t)^\dagger 
\Gamma_{55} = \sum_{xyz} |G(x,y,z,t)|^2,
\end{equation}
and extracted $a m_{PS}$ searching for the large-time plateau in 
the effective mass plot, as a function of $t$. Monitoring the 
behavior of $(a m_{PS})^2$ as a function of $(a m)$, one can 
study the partially conserved axial current and the issues 
related to mass renormalization.

\begin{figure*}
\centerline{
\includegraphics[width=0.41\textwidth]{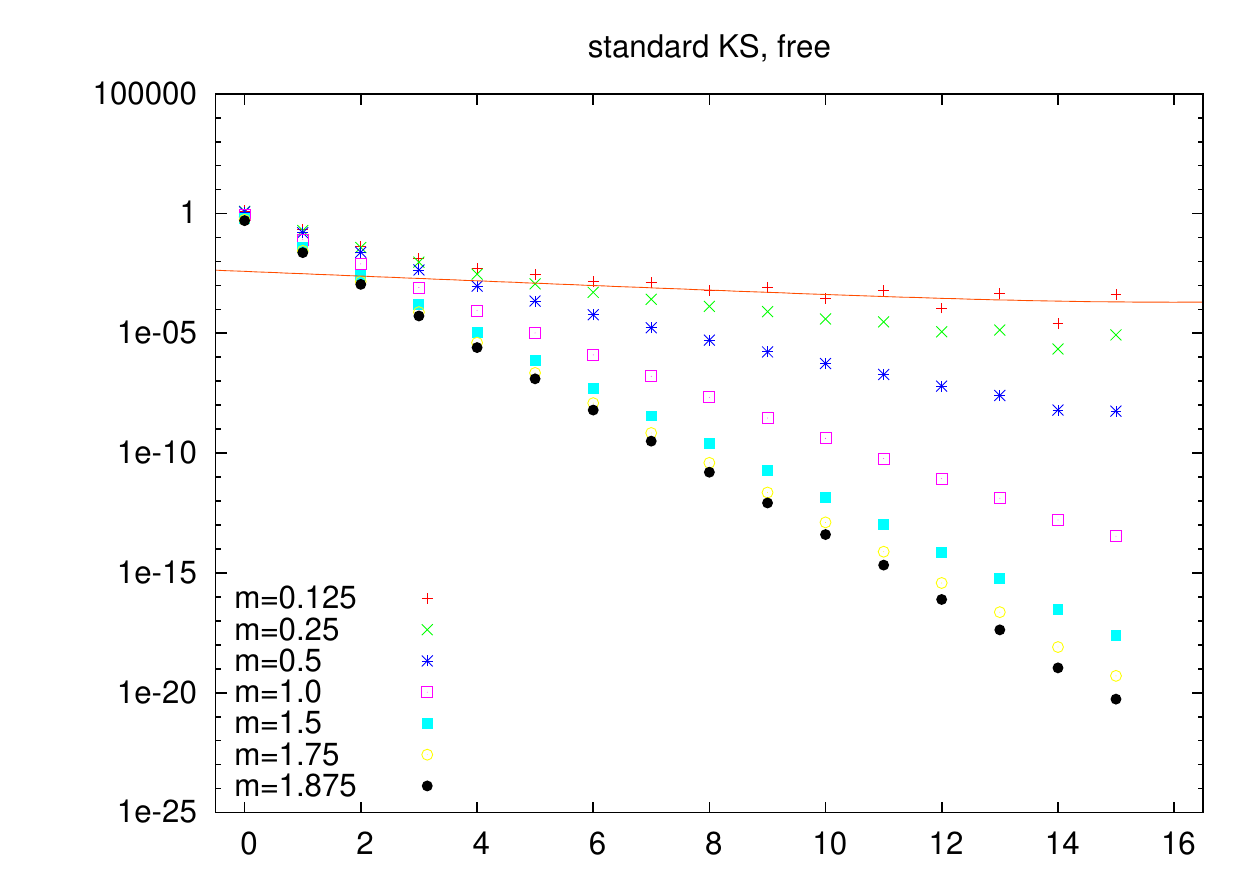} \hfill 
\includegraphics[width=0.41\textwidth]{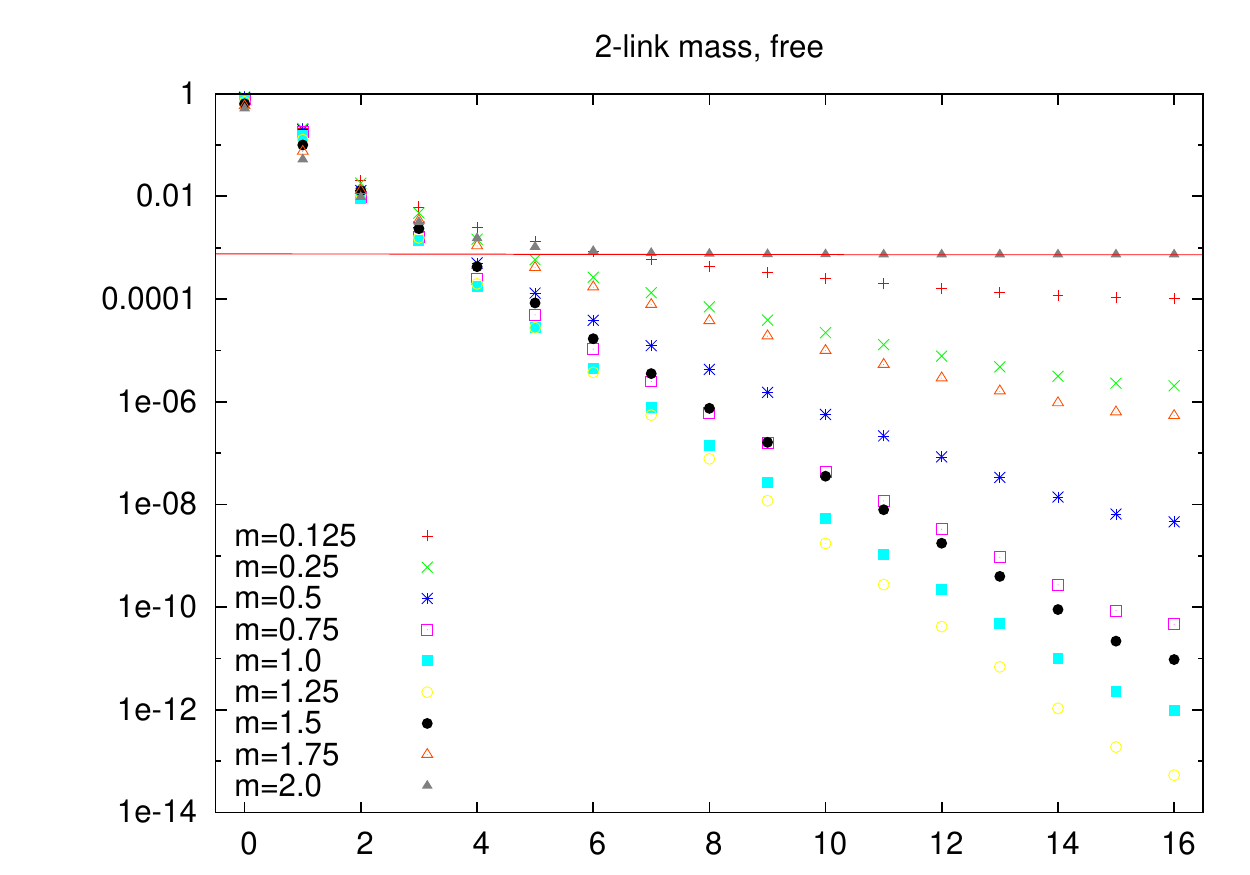} \hfill 
\includegraphics[width=0.41\textwidth]{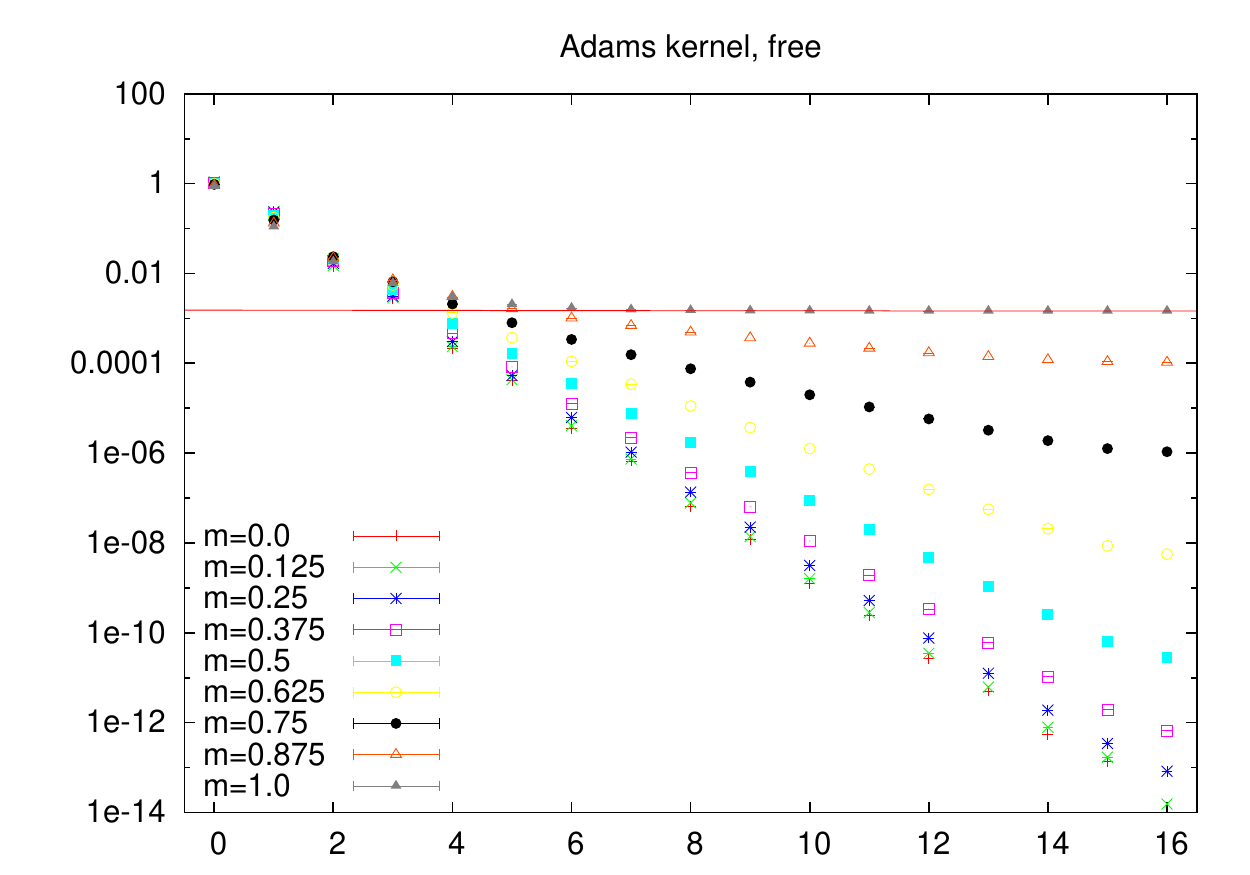} 
}
\caption{The exponential decay of the correlation function 
associated with the lightest meson in the pseudoscalar channel 
for $D_0$ (left), for $D_2$ (center) and $D_4$ (right panel) on a 
free configuration, for different values of the bare quark mass.}
\label{pion_free_conf}
\end{figure*}

Fig.~\ref{pion_free_conf} shows the correlators obtained on a 
free configuration, for $D_0$ (left panel), for $D_2$ (central 
panel) and for $D_4$ (right panel). As expected, the $D_2$ 
operator leads to a massless pion for both $am=0$ and $am=2$.

\begin{figure*}
\centerline{\includegraphics[width=0.58\textwidth]{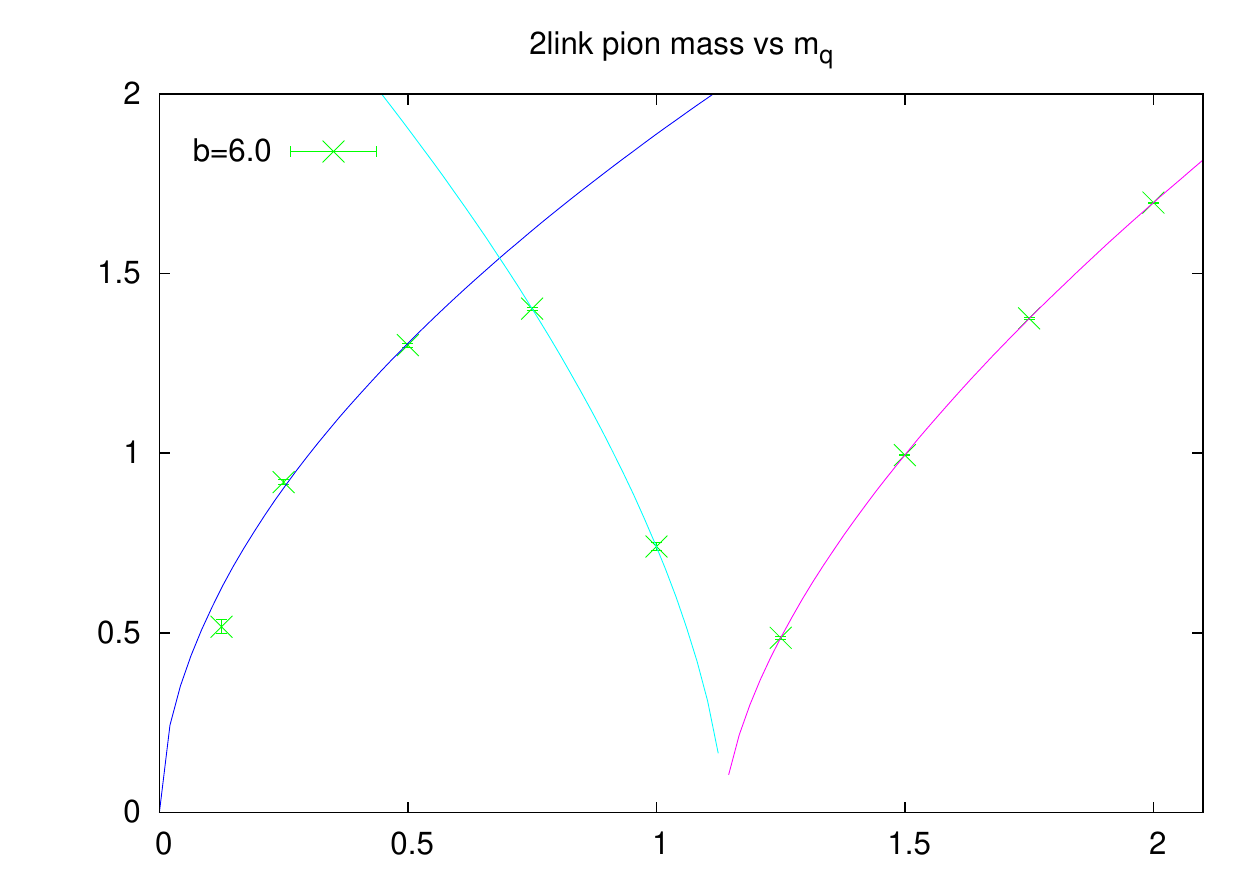} \hfill 
\includegraphics[width=0.58\textwidth]{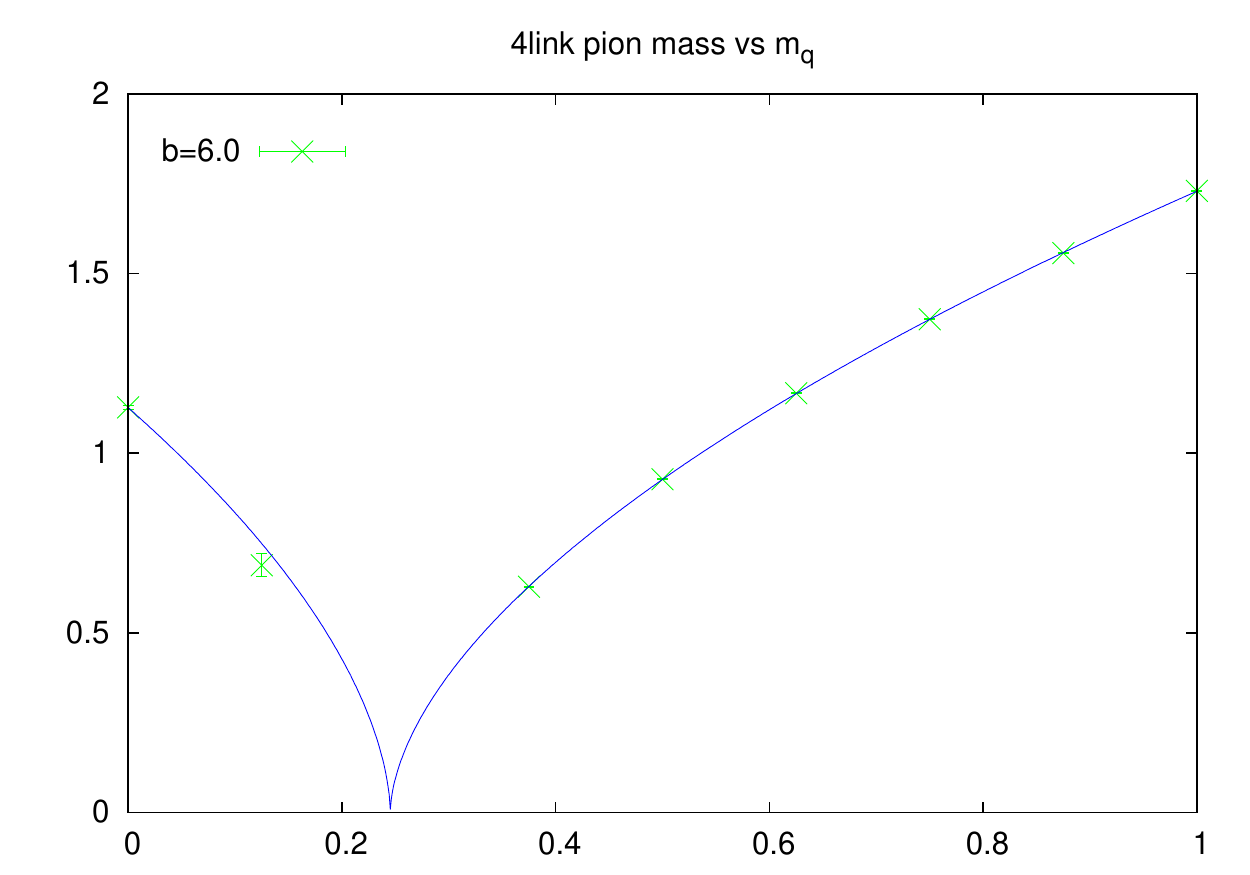}
}
\caption{Tuning of the bare quark mass to obtain a light pion: 
the two plots show the pion mass as a function of the bare quark 
mass, for $D_2$ (left panel) and $D_4$ (right panel) at 
$\beta=6$.}
\label{pion_interacting_conf}
\end{figure*}

For an interacting configuration (at $\beta=6$), the comparison 
between $D_2$ and $D_4$ shown in Fig.~\ref{pion_interacting_conf} 
reveals that for $D_2$ one obtains a massless pion at 
approximately $am\sim 1.15$, while for $D_4$ the same happens for 
$am \sim 0.25$. Comparing these numbers with the values of the 
bare masses corresponding to a massless pseudoscalar state in the 
free limit (1 and 2 respectively), these results give an 
indication that the mass renormalization is more pronounced for 
$D_4$ than for $D_2$. Quantitatively, one can observe that the 
renormalization factor grows exponentially with the length of the 
parallel transporters used: $(0.25/1)^{1/4} \sim (1.15/2)^{1/2}$, 
in agreement with the fact that $D_4$ involves 4-link terms, as 
opposed to $D_2$, in which the mass term is constructed from 2-
link terms.

Remarkably, with the $D_2$ operator, the pion mass shows a 
square-root behaviour of three different kinds: one can approach 
the critical bare quark mass $a m_0 \sim 1.15$ from the left 
or from the right, i.e. from the inside or the outside of the 
$D_2$ spectrum (the behaviour is square-root-like even though the theory 
describes one flavour only -- it is caused by the approach to the Aoki phase). 
In addition, one can also approach the other critical quark 
mass $a m = 0$, corresponding to the central branch of the 
spectrum, which remains zero as in the free case by symmetry of
the average spectrum. The transition from one branch to another seems rather 
abrupt, and the scaling of the pion mass can be observed over a 
broad range of quark masses approaching zero.

The lesson is that $D_2$ may provide a cost-effective way to 
simulate $N_f=2$ light quark species, {\em without} fine-tuning 
of the bare quark mass to approach the chiral limit.

\section{Conclusions}
\label{sec:conclusions}
In this work, we performed a numerical study of staggered 
Dirac operators with a taste-dependent mass term. We restricted 
our attention to operators including mass terms with tensor or 
pseudoscalar structure in taste space: their $\gamma_5$-hermiticity 
properties are such, that their eigenvalues come in 
complex conjugate pairs (as is the case for the usual staggered 
Dirac operator), leading to a real fermionic determinant, which
is non-negative in the absence of negative real eigenvalues. 

Such operators were proposed by 
Adams~\cite{staggeredoverlap1,staggeredoverlap2} and by 
Hoelbling~\cite{Hoelbling:2010jw}. We compared their properties 
both in the free limit and on interacting configurations at 
typical values of the gauge coupling.

Our results show that these operators can indeed be used to 
separate the low-lying modes and reduce the number of tastes, in 
a way characterized by well-defined topological
properties. Our study of the spectral flow reveals that, for the 
4-link operator (with a taste-pseudoscalar mass term), the gap in 
the eigenvalue spectrum tends to close rather early, obstructing 
an easy identification of the eigenvalue crossings, which are 
related to the index. As one might have expected, the 2-link 
operator shows markedly more robustness to gauge fluctuations.

We also performed an elementary study of pion propagators, 
which shows that the lightest meson is rather easy to isolate 
without explicitly disentangling spin and taste degrees of 
freedom. Approaching the chiral limit requires in general the 
fine-tuning of an additive mass term, as for Wilson fermions. One 
important exception occurs for the 2-link operator: if one 
chooses the middle branch of the spectrum, one can study a theory 
with two tastes, where the additive mass renormalization vanishes 
due to the symmetry of the spectrum. Therefore, no fine-tuning is 
needed.

Although the 2-link operator was designed to produce a 
single taste (with a fine-tuned additive mass), it may well be 
that its most promising use is to simulate two tastes without 
additive mass renormalization.
Note that the heavy doubler modes do not completely decouple
in that situation. In the background of a topological charge $Q$,
they contribute real eigenvalues $\sim (+1/a)^Q$ and $(-1/a)^Q$,
making the determinant negative when $Q$ is odd. The $\theta$-parameter
is thus equal to $\pi$. This sign $(-1)^Q$ should be removed by hand
(or simply ignored) in order to simulate the $\theta=0$ theory.

Finally, we studied the properties of an overlap operator 
with a $D_4$ kernel (see Appendix). We found that its locality properties are 
similar to those of the operator based on a Wilson kernel. As it 
concerns the computational cost for a quark propagator 
calculation, we found that, in the free limit or on very smooth 
gauge configurations, the inversion of the operator based on a 
kernel with a four-link mass term is almost one order of 
magnitude faster than using an overlap with Wilson kernel. 
However, we also observed a significant loss of efficiency on 
interacting (quenched) configurations at $\beta=6$, where the 
operator with the $D_4$ kernel is only approximately twice as 
fast as that with a Wilson kernel. The reason for this can 
probably be traced back to the fact that the four-link 
transporters in the mass term are more sensitive to the effect of 
the fluctuations in gauge configurations on coarser lattices.  
Our crude assessment indicates that this new, staggered, overlap 
operator does not bring a major computational advantage over a 
Wilson kernel, while producing two degenerate flavors, but 
without the full $SU(2)$ flavor symmetry.

Two copies of an overlap operator with a kernel based on 
Hoelbling's 2-link operator would give more flexibility, e.g. 
that of simulating two flavors with unequal masses, for a similar 
computer effort.

{\bf Note added:} 
After this paper was completed, a difficulty with the Hoelbling
operator $D_2$ eq.(\ref{twolinkkernel}) was pointed out by Steve Sharpe,
and clarified by David Adams, during the Yukawa Institute Workshop
``New Types of Fermions on the Lattice''. It appears that the Hoelbling
operator lacks sufficient rotational symmetry, so that fine-tuned
Wilson loop counterterms will presumably be needed to maintain hypercubic
rotational symmetry in unquenched simulations. Adams' operator $D_4$
eq.(\ref{fourlinkkernel}) does not suffer from this problem.

\section*{Acknowledgements}
This research was supported by the Natural Sciences and 
Engineering Research Council of Canada, by the Academy of 
Finland, project 1134018, and in part by the National Science 
Foundation under Grant No. PHY11-25915. Ph.~de~F. thanks the 
Yukawa Institute for Theoretical Physics, Kyoto, Japan, for 
hospitality. Ph.~de~F. and M.~P. gratefully acknowledge the Kavli 
Institute for Theoretical Physics in Santa Barbara, USA, for 
support and hospitality during the ``Novel Numerical Methods for 
Strongly Coupled Quantum Field Theory and Quantum Gravity'' 
program, during which part of this work was done. We thank 
D.~H.~Adams, M.~Creutz, S.~D\"urr, C.~Hoelbling, S.~Kim, T.~Kimura, T.~Misumi, 
A.~Ohnishi, S.~Sharpe and all participants of the Yukawa Institute Workshop
``New Types of Fermions on the Lattice'' for discussions.


\appendix{}
\section{Staggered overlap operator}
\label{appendix}
\renewcommand{\theequation}{A.\arabic{equation}}
\setcounter{equation}{0}

We also studied the properties of an overlap operator based 
on a staggered $D_4$ kernel, as originally proposed in \cite{staggeredoverlap1}. The construction is completely standard:
\begin{equation}
\label{overlap}
\Dov = \frac{\rho}{a} \left( 1+ \frac{D_4}{\sqrt{D_4^\dagger D_4}} \right) 
\end{equation}
and leads to two exactly massless physical fermions in the 
continuum limit, without fine-tuning. As compared to a 
conventional overlap operator based on a Wilson kernel, the 
potential advantages of this construction are related to the 
reduced kernel size ($D_4$ is a matrix of size four times smaller 
than a Wilson operator on the same lattice).
We take $\rho=1$.

To understand the effectiveness of an overlap operator with a 
$D_4$ kernel, the first important issue to be discussed is the 
locality of the operator. As is well-known, an overlap operator 
is not ultra-local~\cite{Horvath:1998cm}. Its locality properties 
can be studied by looking at the decay of its matrix elements 
between source and sink at sites $x$ and $y$ (which we denote as 
$M_{x,y}$, where, for simplicity, we only show the indices 
corresponding to the site coordinates), as a function of the 
distance between $x$ and $y$~\cite{Hernandez:1998et}. To this 
end, in the two plots at the top of Fig.~8 we show
the decay of $|M_{x,y}|$ against $|x-y|_1$, the 1-norm distance 
(or ``Manhattan distance'') between the sites $x$ and $y$, 
comparing the matrix elements of an overlap operator obtained 
using a $D_4$ kernel (left panel) or a conventional Wilson kernel 
(right panel). Although the $D_4$ kernel is less local than the 
Wilson kernel, the locality properties of the corresponding 
overlap operators are comparable. This appears quite clearly in 
the plots displayed at the bottom of the figure, in which the 
results for the two operators are shown together, for a cold 
configuration (left panel) and for a configuration at $\beta=6$ 
(right panel).

\begin{figure*}
\centerline{
\includegraphics[width=0.50\textwidth]{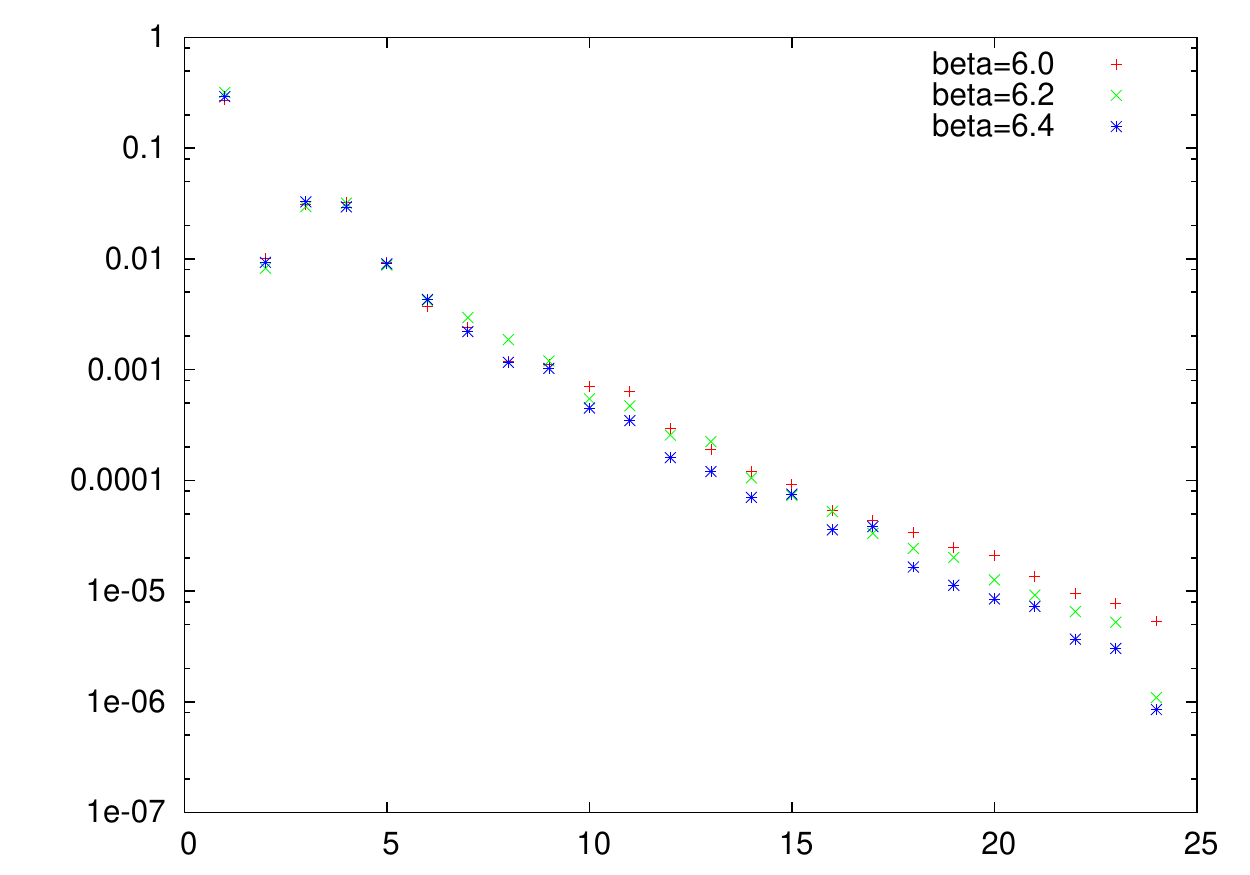} \hfill 
\includegraphics[width=0.50\textwidth]{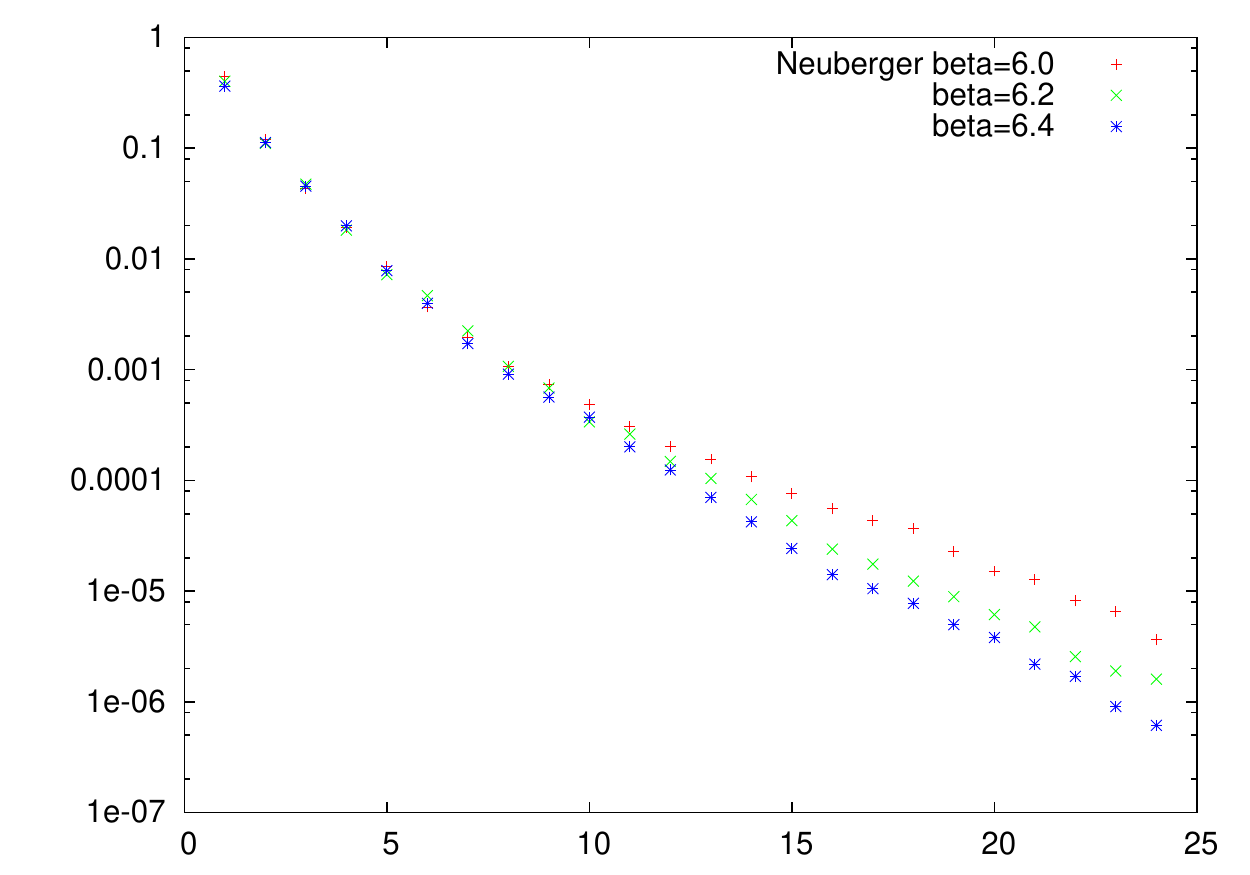}
}
\centerline{
\includegraphics[width=0.50\textwidth]{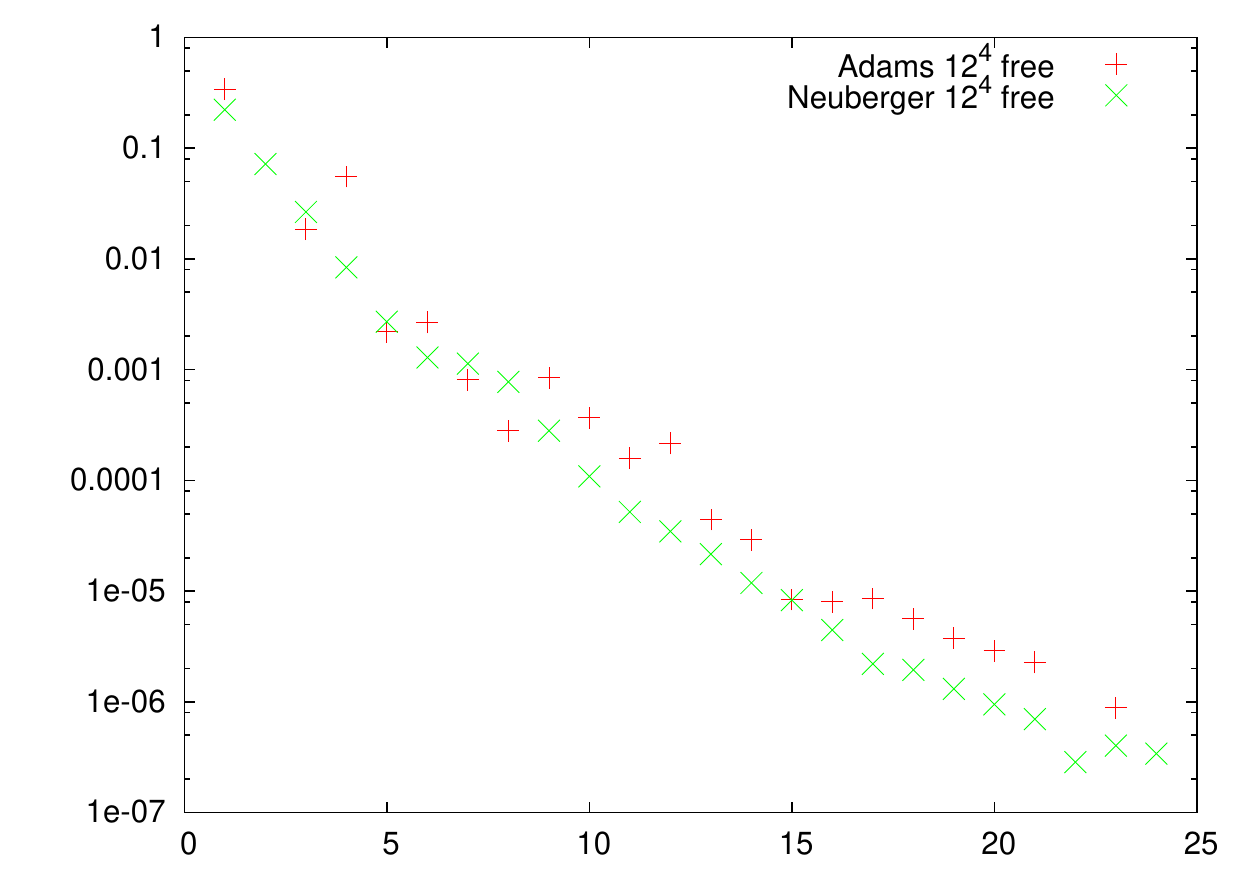} \hfill 
\includegraphics[width=0.50\textwidth]{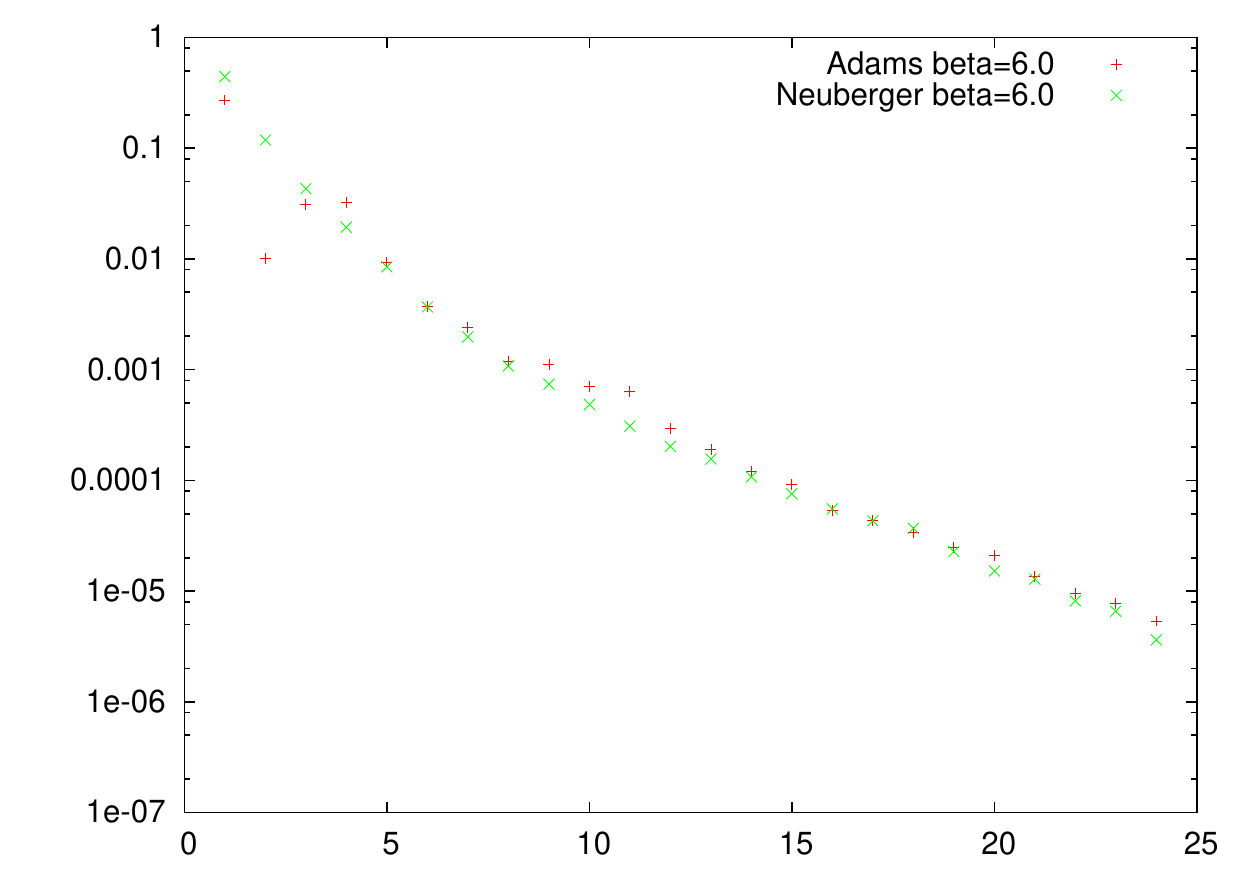}
}

\label{locality}
\caption{The plots at the top of the figure show the decay of the 
matrix elements of an overlap operator obtained using a $D_4$ 
kernel (left) or a Wilson kernel (right), as a function of the 1-
norm distance between the sites. The results from the two
operators are displayed together in the two bottom plots, for a 
cold (bottom left panel) and a $\beta=6$ configuration (bottom 
right panel).}
\end{figure*}

Another important factor in the efficiency of a lattice Dirac 
operator is the cost of applying the operator to a vector. The multiplication 
by the kernel is about twice as fast, if one uses $D_4$ instead 
of a Wilson kernel (staggered fields do not have an explicit 
spinor index but $D_4$ has twice as many non-zero elements as the 
Wilson operator). In the computation of the sign of $\tilde{H}=\Gamma_{55} D_4$, 
using the conjugate gradient (CG) method, and no deflation, the 
gain with respect to a conventional Wilson kernel is a factor 
ranging from approximately 2-3 to about 8. However, these numbers 
are only gross estimates, and could be improved, e.g., by 
optimizing the parameters of the $D_4$ kernel. Similarly, an 
improved form for the kinetic operator, link smearing (for the 
kinetic and/or the mass term), and standard tricks related to 
deflation, preconditioning, etc... could be applied.

To discuss the computational cost of the inversion of the 
operator, we compared the two overlap operators on the same pure-glue, $\beta=6$, background on a $12^4$ lattice, using the same, 
basic, inner/outer CG algorithm. In our computation, we evaluated 
the propagator as the solution of the equation:
\begin{equation}
(D_{ov}+m)^\dagger (D_{ov}+m) x = (D_{ov}+m)^\dagger b
\end{equation}
with $ma=0.1$,
using a conjugate gradient (CG) iterative solver: at each 
iteration, ${\rm sign}(H)$ is applied to a vector $v$ through a 
2-pass Lanczos process. One builds a tridiagonal matrix $T$ and 
takes the signs of its eigenvalues (which are representative of 
those of $H$), then reconstructs ${\rm sign}(H) v$, as described 
in ref.~\cite{Borici:1998mr}. The results are displayed at the 
top of  Fig.~8: the three plots (from 
left to right) show the computational cost for the outer CG 
iteration, for the matrix-times-vector multiplication, and the 
total CPU cost. For comparison, we also show the analogous 
results in the free-field case, in the plots at the bottom of the 
figure. This comparison shows that the computational advantages 
expected from elementary arguments, and observed in the free 
limit, turn out to be dramatically reduced in ``realistic'' 
interacting configurations. Again, this reduction points to a 
reduction in the eigenvalue gap of $D_4$.

\begin{figure*}
\centerline{ 
\includegraphics[width=0.39\textwidth]{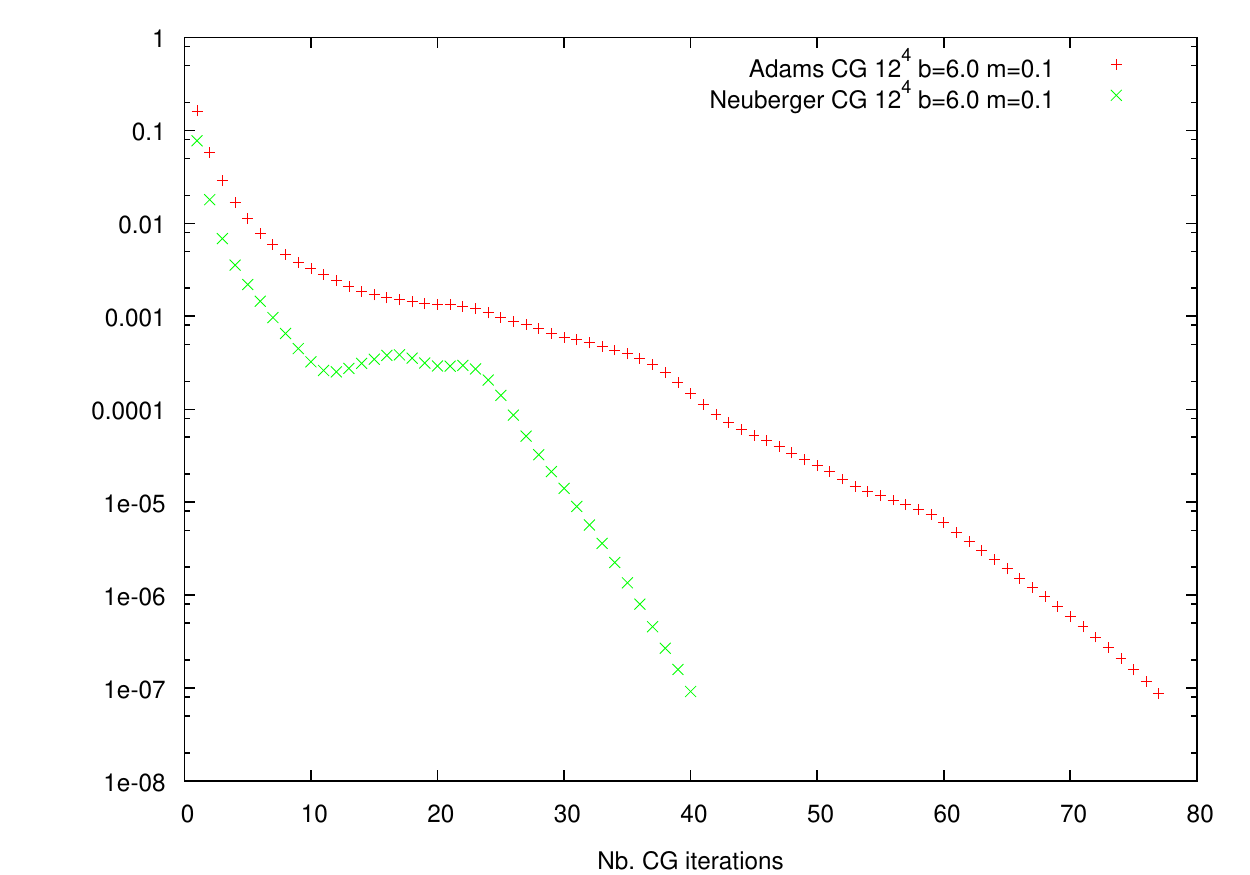} \hfill 
\includegraphics[width=0.39\textwidth]{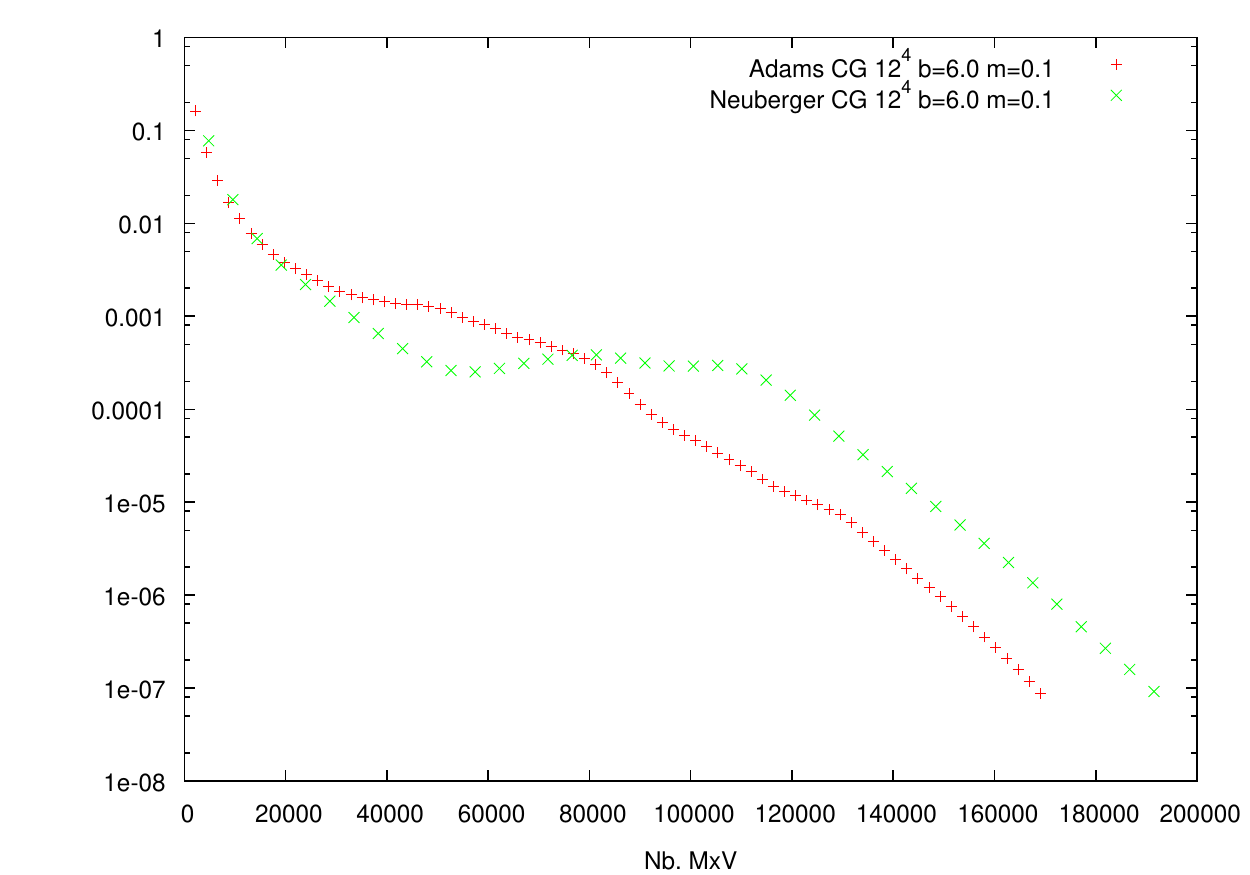} \hfill 
\includegraphics[width=0.39\textwidth]{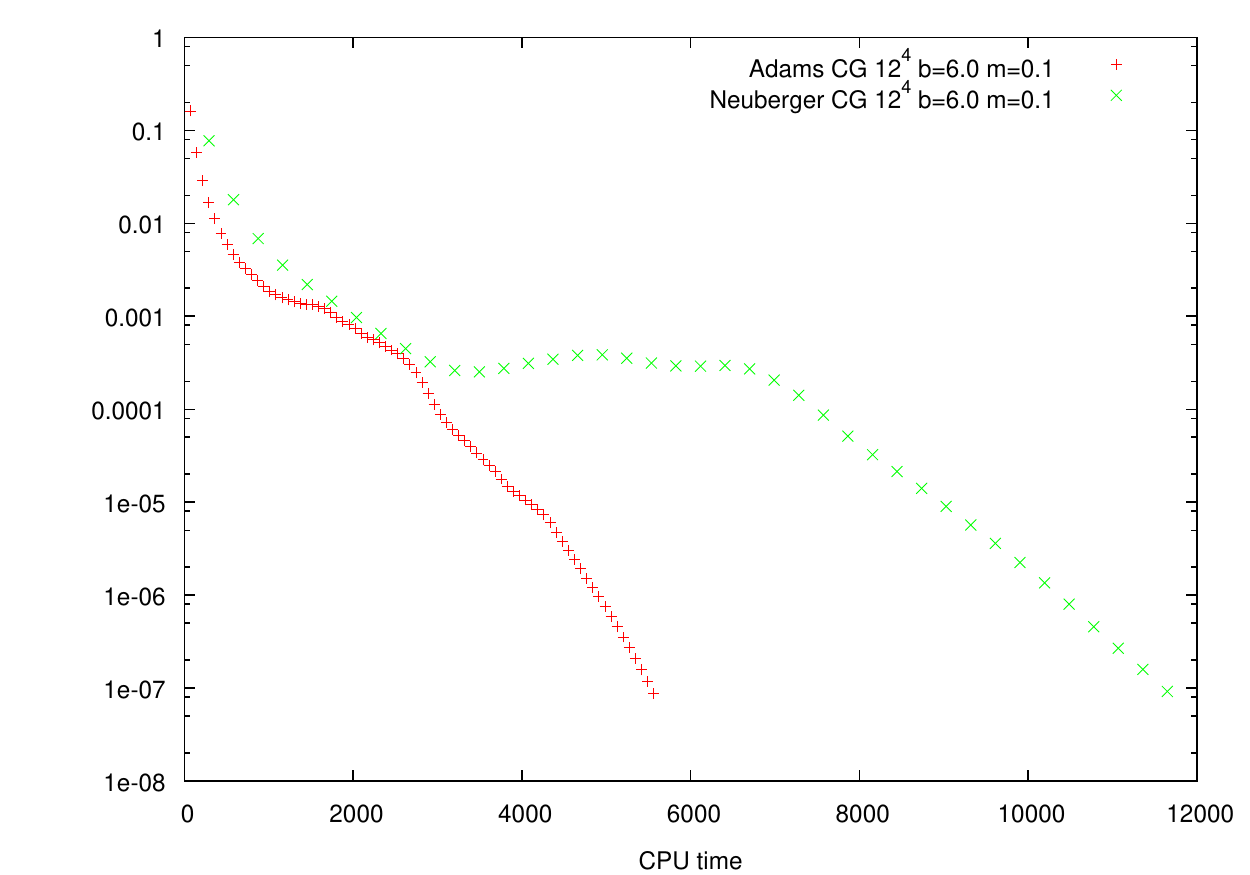}
}
\centerline{ 
\includegraphics[width=0.39\textwidth]{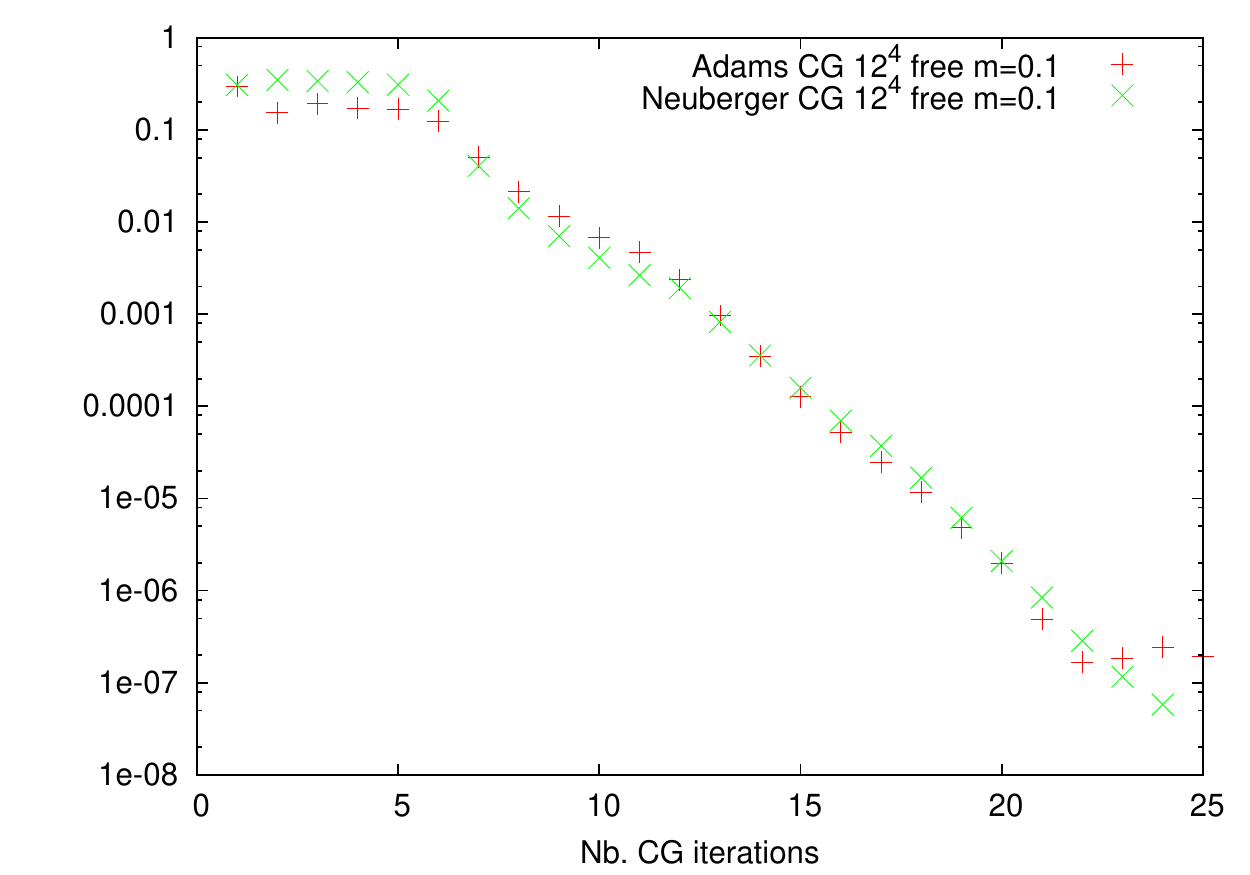} \hfill 
\includegraphics[width=0.39\textwidth]{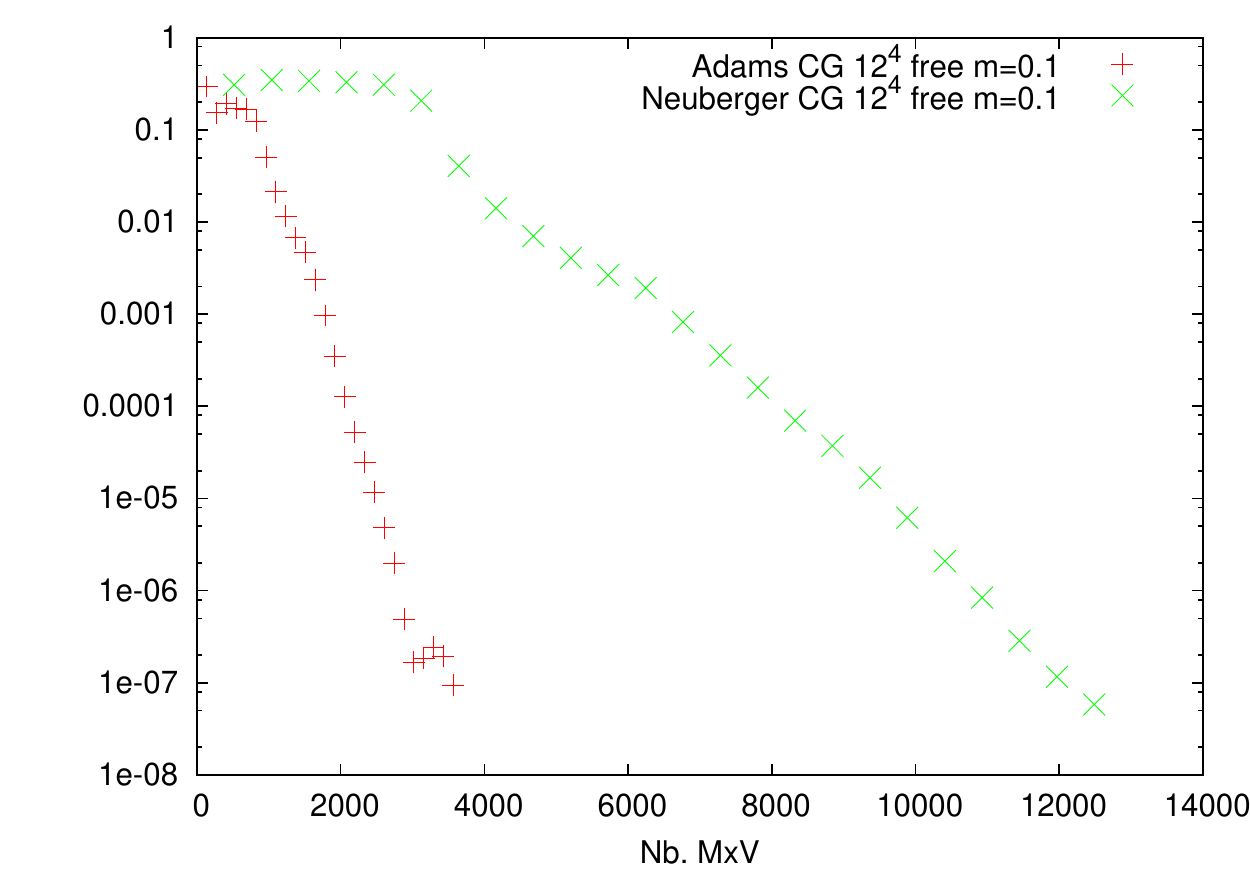} \hfill 
\includegraphics[width=0.39\textwidth]{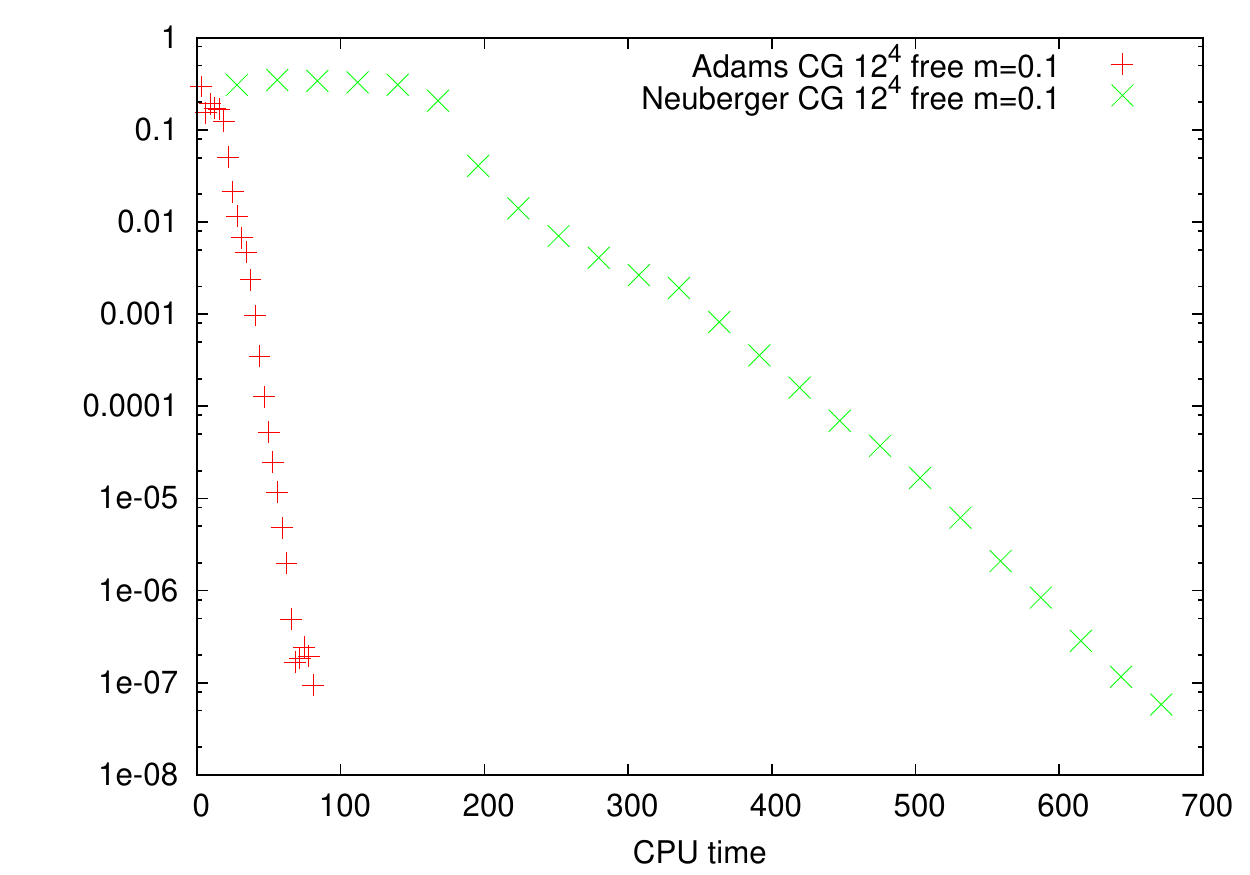}
}
\label{inversion_comparison}
\caption{Top row: Comparison of the computational costs for the 
overlap operator inversion at a given level of precision, using a 
$D_4$ (red points) or a Wilson (green symbols) kernel: the left 
panel displays the costs related to the outer CG iteration, the 
central plot shows the costs of the matrix-times-vector 
multiplication, and finally the right panel displays the total 
CPU cost. For comparison, the plots in the bottom row show the 
analogous results for a free configuration.}
\end{figure*}

\end{document}